\documentclass{aa}
\usepackage{graphicx}
\begin{document}
\title{Spectroscopic observations of the \\candidate sgB[e]/X-ray binary
\object{CI~Cam}}
   \author{R. I. Hynes\inst{1} \and
           J. S. Clark\inst{2} \and 
           E. A. Barsukova\inst{3} \and 
	   P. J. Callanan\inst{4} \and
           P. A. Charles\inst{1} \and
	   A. Collier Cameron\inst{5} \and \\
           S. N. Fabrika\inst{3} \and
	   M. R. Garcia\inst{6} \and
           C. A. Haswell\inst{7}\fnmsep\thanks{Guest Observer, 
           McDonald Observatory, University of Texas at Austin.} \and
	   Keith Horne\inst{5} \and
           A. Miroshnichenko\inst{8,9} \and
	   I. Negueruela\inst{10} \and \\
	   P. Reig\inst{11,12} \and
           W. F. Welsh\inst{13}\fnmsep$^{\star}$\and
	   D. K. Witherick\inst{2}
   }
   \authorrunning{R. I. Hynes et al.}
   \titlerunning{Spectroscopic observations of \object{CI~Cam}}

   \offprints{R. I. Hynes, \email{rih@astro.soton.ac.uk}}

   \institute{Department of Physics and Astronomy, University of Southampton, 
              Southampton, SO17 1BJ, UK \and
              Department of Physics and Astronomy, University College
              London, Gower Street, London, WC1E 6BT, UK \and
              Special Astrophysical Observatory, 
              Nizhnij Arkhyz, 369167, Russia \and 
              University College, Department of Physics, Cork, Ireland \and
	      School of Physics and Astronomy, University of St. Andrews, 
              North Haugh, St. Andrews, Fife KY16 9SS, UK \and
	      Harvard Smithsonian Center for Astrophysics, 60 Garden
              Street, Cambridge, MA 02138, USA\and
              Department of Physics and Astronomy, The Open
              University, Walton Hall, Milton Keynes, MK7 6AA, UK \and 
	      University of Toledo, Dept. of Physics and Astronomy, 
              Toledo, OH 43606, USA \and
              Pulkovo Observatory, 196140 Saint-Petersburg, Russia\and
              Observatoire de Strasbourg,11 rue de l'Universite, 
              67000 Strasbourg, France \and
              Physics Department, University of Crete, P.O. Box 2208, GR-71003,
              Heraklion, Greece \and
              Foundation for Research and Technology--Hellas, GR-71110,
              Heraklion, Greece \and
	      Dept.\ of Astronomy, San Diego State University, San
              Diego, CA 92182, USA}

\date{Received 7 March 2002; accepted 1 July 2002}

\abstract{We present a compilation of spectroscopic observations of
the sgB[e] star \object{CI~Cam}, the optical counterpart of
\object{XTE~J0421+560}.  This includes data from before, during, and
after its 1998 outburst, with quantitative results spanning 37 years.
The object shows a rich emission line spectrum originating from
circumstellar material, rendering it difficult to determine the nature
of either star involved or the cause of the outburst.  We collate all
available pre-outburst data to determine the state of the system
before this occurred and provide a baseline for comparison with
outburst and post-outburst data.  During the outburst all lines become
stronger, and hydrogen and helium lines become significantly broader
and asymmetric.  After the outburst, spectral changes persist for at
least three years, with Fe\,\textsc{ii} and [N\,\textsc{ii}] lines
still a factor of $\sim2$ above the pre-outburst level and
He\,\textsc{i}, He\,\textsc{ii}, and N\,\textsc{ii} lines suppressed
by a factor of 2--10.  We find that the spectral properties of
\object{CI~Cam} are similar to other sgB[e] stars and therefore
suggest that the geometry of the circumstellar material is similar to
that proposed for the other objects: a two component outflow, with a
fast, hot, rarefied polar wind indistinguishable from that of a normal
supergiant and a dense, cooler equatorial outflow with a much lower
velocity.  Based on a comparison of the properties of \object{CI~Cam}
with the other sgB[e] stars we suggest that \object{CI~Cam} is among
the hotter members of the class and is viewed nearly pole-on.  The
nature of the compact object and the mechanism for the outburst remain
uncertain, although it is likely that the compact object is a black
hole or neutron star, and that the outburst was precipitated by its
passage through the equatorial material.  We suggest that this
prompted a burst of supercritical accretion resulting in ejection of
much of the material, which was later seen as an expanding radio
remnant.  The enhanced outburst emission most likely originated either
directly from this supercritical accretion, or from the interaction of
the expanding remnant with the equatorial material, or from a
combination of both mechanisms.
\keywords{stars:individual(CI~Cam) -- stars:emission line, B[e] --
stars:radio emission stars:CI~Cam -- binaries: close}}
\maketitle
\section{Introduction}
\label{Intro}

On 1998 April 2 Smith et al.\ (\cite{Smith:1998a}) reported an {\it
RXTE} All-Sky Monitor (ASM) detection of a bright, rapidly rising
X-ray transient designated \object{XTE~J0421+560}.  Subsequently,
Marshall et al.\ (\cite{Marshall:1998a}) used the {\it RXTE}
Proportional Counting Array (PCA) to refine the best fit position with
an error circle of 1\,arcmin radius. The bright ($V\sim11$) B[e] star
\object{CI~Cam}(=\object{MWC~84}) was found to lie near to the centre
of this error circle.  Spectroscopic observations by Wagner et al.\
(\cite{Wagner:1998a}) on 1998 April 3 revealed a rich emission line
spectrum, similar to that reported by Downes (\cite{Downes:1984a}; see
Sect.~\ref{QuiescentSpec}), but with the presence of He\,\textsc{ii}
emission features. These features had not been reported in previous
spectra, and so by analogy to other X-ray binaries Wagner et al.\
(\cite{Wagner:1998a}) proposed it to be the optical counterpart of
\object{XTE\,J0421+560}. Photometric observations of the source at
this time (e.g.\ Robinson et al.\ \cite{Robinson:1998a}, Garcia et
al.\ \cite{Garcia:1998a}, Hynes et al.\ \cite{Hynes:1998a}) showed
that \object{CI~Cam} was some 2--3\,mag brighter than had previously
been reported.

Hjellming \& Mioduszewski (\cite{Hjellming:1998a}) reported the
detection of a transient 19\,mJy radio source at 1.4\,GHz,
corresponding to the optical position of \object{CI~Cam} on 1998 April
1, thus confirming the identification of \object{CI~Cam} as the
optical counterpart.  Subsequent observation of rapid radio
variability established that the radio emission was of non-thermal
(synchrotron) origin (Hjellming \& Mioduszewski
\cite{Hjellming:1998b}).  Long term observations indicate that after
the initial flare the radio emission underwent an unusually slow
decay, with a 15\,GHz flux of $\sim1.5$\,mJy about 40 months after the
initial outburst (Pooley, priv.\ comm.).  High spatial resolution maps
obtained after the outburst indicated the presence of a clumpy
ejection nebula (Mioduszewski et al., in preparation).  These ejecta
expand at $\sim 1.0-1.5$\,mas\,d$^{-1}$, corresponding to an expansion
velocity $\sim5000$\,km\,s$^{-1}$ for a distance of 5\,kpc.

Given the distance estimates (and optical luminosity implied) for
\object{CI~Cam} (e.g.\ $\log L/L_{\odot} \geq4.86$; Clark et al.\
\cite{Clark:2000a}; Robinson, Ivans \& Welsh \cite{Robinson:2002a}) it
is clear that if it is a binary, as is likely, it is a high mass X-ray
binary (HMXB). However \object{CI~Cam} does not sit comfortably within
the traditional divisions of HMXB mass donors into classical Be stars
($\sim70$\,per cent) and OB supergiants ($\sim30$\,per cent).  While
its luminosity suggests that it belongs to the later subset, such
systems are typically short period binaries which accrete via Roche
lobe overflow or direct wind fed accretion producing {\em persistent}
X-ray emission (typically modulated at the orbital period).  The
presence of a rich emission line spectrum including forbidden lines,
and a near IR excess due to hot dust (the observational criteria for
the B[e] phenomenon; Allen \& Swings \cite{Allen:1976a}; Lamers et
al.\ \cite{Lamers:1998a}) also mark a distinction from the other
supergiant HMXB systems.  Among the stars showing the B[e] phenomenon,
the high luminosity of \object{CI~Cam} makes it a Galactic counterpart
to the Magellanic Cloud supergiant B[e] stars (sgB[e] stars) and we
will refer to it as such for the rest of this work.  \object{CI~Cam}
therefore appears to be the first bona fide sgB[e] star HMXB known,
although direct evidence for binarity has proven elusive and a chance
encounter of a compact object with the sgB[e] star although highly
unlikely, cannot be ruled out.

This work presents a compilation of spectroscopy obtained before,
during and after the 1998 outburst.  A companion photometric
compilation has been presented by Clark et al.\ (\cite{Clark:2000a}).
Some of the outburst data included here has previously been presented
by Barsukova et al.\ (\cite{Barsukova:1998a}) and Barsukova et al.\
(\cite{Barsukova:2002a}).  In Sect.~\ref{QuiescentSpec} we summarise
the available pre-outburst data, including archival data with
quantitative spectroscopy spanning $\sim30$\,years before the X-ray
outburst and additional unpublished pre-outburst spectra.  From these
we identify the typical pre-outburst strengths of spectral lines and
discuss their stability.  We then in Sect.~\ref{OutburstSpec} describe
a series of new spectra running from a few days after the X-ray
outburst to $\sim3$\,years later.  In Sect.~\ref{ExtinctionSection} we
discuss extinction and distance estimates for the system and in
Sect.~\ref{CompanionSection} we review what is known about the mass
donor star.  Sect.~\ref{FluxSection} examines the changes in the
continuum flux distribution and Sect.~\ref{SpecEvol} the spectral
lines and how these evolve through the outburst.  Sect.~\ref{Rapid}
tests for the presence of shorter timescale variability.  Finally in
Sect.~\ref{Discussion} we will discuss how all of these clues can help
us build a picture of the nature of the system and the outburst
mechanism and in Sect.~\ref{conc} we summarise our conclusions.
%
%%%%%%%%%%%%%%%%%%%%%%%%%%%%%%%%%%%%%%%%%%%%%%%%%%%%%%%%%%%%%%%%%%%%%%%%%%%%%%%
%
\section{Pre-outburst observations}
\label{QuiescentSpec}
Since \object{CI~Cam} is bright and has a strong emission line
spectrum, there exists a relatively large data set dating from 1933 to
the present day.  We present a summary of these observations, which
include previously unpublished optical spectra from 1987 and 1994, to
quantify an average `quiescent' state for \object{CI~Cam}.
Table~\ref{QuiescentEWTable} summarises the pre-outburst equivalent
widths.

\subsection{Published archival spectroscopy}
\label{ArchivalSpec}
The first observations of \object{CI~Cam} were made by Merrill \&
Burwell (\cite{Merrill:1933a}) and showed strong H\,\textsc{i} Balmer
and weak Fe\,\textsc{ii} emission. He\,\textsc{i} was seen to be
strongly in emission but He\,\textsc{ii} was absent.  The first
quantitative spectroscopic observations of \object{CI~Cam} date from
Chkhikvadze (\cite{Chkhikvadze:1970a}). Spectra obtained during the
period 1964--67 show strong H\,\textsc{i} and He\,\textsc{i} emission
lines with no evidence of He\,\textsc{ii} emission. Fe\,\textsc{ii}
lines were also in emission, the strongest being multiplets 27, 28, 37
and 38; there was no evidence for the Mg\,\textsc{ii} 4481\,\AA\ line.
Based on the strength of the Balmer emission lines Chkhikvadze
(\cite{Chkhikvadze:1970a}) suggested a spectral class of O6--B0 for
\object{CI~Cam}.

Allen \& Swings (\cite{Allen:1976a}) describe further spectroscopic
observations of \object{CI~Cam} (although no date for the observations
is provided). They note numerous emission lines of He\,\textsc{i},
Fe\,\textsc{ii} and Si\,\textsc{ii}.  [N\,\textsc{ii}] lines at 5755
and 6584\,\AA\ were present (the latter in the wings of the very
strong H$\alpha$ line); [O\,\textsc{i}] lines at 6300 \& 6363\,\AA\
are possibly also detected. An estimate for the density of the
circumstellar envelope of $N_{\rm e} >10^{6} - 10^7$\,cm$^{-3}$ was
derived from the strength of the [N\,\textsc{ii}] lines. Higher
excitation lines such as [O\,\textsc{iii}] and [N\,\textsc{iii}]
appeared to be absent.

A 4000--7000\AA\ spectrum was obtained by Downes (\cite{Downes:1984a})
in 1984 January and once again was dominated by strong H\,\textsc{i},
He\,\textsc{i} and Fe\,\textsc{ii} lines.

Miroshnichenko (\cite{Miroshnichenko:1995b}) describes observations
made between 1986 September and 1987 December.  The spectrum was
dominated by strong H\,\textsc{i} and He\,\textsc{i} lines, with
numerous weak Fe\,\textsc{ii} lines also present (as was
C\,\textsc{ii} emission at 4267 \& 7234\AA). The H\,\textsc{i} Balmer
lines were symmetrical and single peaked, with wings extending to
$\pm$250\,km\,s$^{-1}$ for H$\alpha$ and H$\beta$.

Finally, Jaschek \& Andrillat (\cite{Jaschek:2000a}) report
observations from 1992 and 1998, the latter only two months before the
outburst.  450 emission lines were observed, $\sim55$\,percent from
Fe\,\textsc{ii}, the remainder including H\,\textsc{i},
He\,\textsc{i}, O\,\textsc{i}, N\,\textsc{i}, Si\,\textsc{ii},
Mg\,\textsc{ii}, [O\,\textsc{i}], [N\,\textsc{ii}], and
[Fe\,\textsc{ii}].  The Balmer lines show a very steep Balmer
decrement.  O\,\textsc{i} 8446\,\AA\ is unusually strong, likely due
to Ly$\beta$ fluorescence.  Very few differences between the 1992 and
1998 epochs were seen.  Since the latter spectra represent the highest
quality pre-outburst spectra, as well as the closest to the X-ray
outburst, we reanalysed them to establish a quantitative pre-outburst
baseline.
\subsection{Unpublished spectra taken at the Special Astrophysical Observatory}
\label{PreoutburstSAOSection}
Three further unpublished observations of \object{CI~Cam} were
obtained by A. Miroshnichenko at the 6\,m BTA telescope of the Special
Astrophysical Observatory of the Russian Academy of Sciences (SAO RAS)
on 1987 April 5 ($\sim$4460--5000\AA ) and 1994 January 21
($\sim$4000--4950 \& $\sim$5700--6800\AA ).

The observations were made with a medium-resolution spectrograph
SP-124 and a 1024-element one-dimensional photo-electric detector
(Drabek et al.\ \cite{Drabek:1986a}).  Data reduction was performed
with the {\sc sipran} software developed at SAO RAS; no attempt at
flux calibration was made.  Due to the poor dynamic range of the
scanner it was possible to obtain either profiles from the strong
lines on an underexposed continuum, or a properly exposed continuum
with saturated strong lines. Of the data presented here, the 1987
spectrum was exposed to search for weak lines, while the two spectra
obtained in 1994 were obtained to determine the line profiles of the
stronger emission lines.  Consequently the continuum (and weaker
lines) were poorly exposed for these spectra.  We have chosen to
include an equivalent width for the lines identified from the
5700-6700\AA\ spectrum, although we caution that the continuum level
was uncertain in this spectrum.

\subsection{Archival spectra from Haute-Provence}
\label{OHPReductionSection}
We obtained the 1998 spectra reported by Jaschek \& Andrillat
(\cite{Jaschek:2000a}) from the archive of the Observatoire de
Haute-Provence (OHP).  These spectra were obtained with the Aurelie
spectrograph on the 1.52\,m telescope using a Thomson array of 2026
photodiodes and a 300 line\,mm$^{-1}$ grating, yielding a resolution
of about 1.3\,\AA.

\subsection{The `average' quiescent spectrum}
\begin{figure*}
\resizebox{\hsize}{!}{\includegraphics{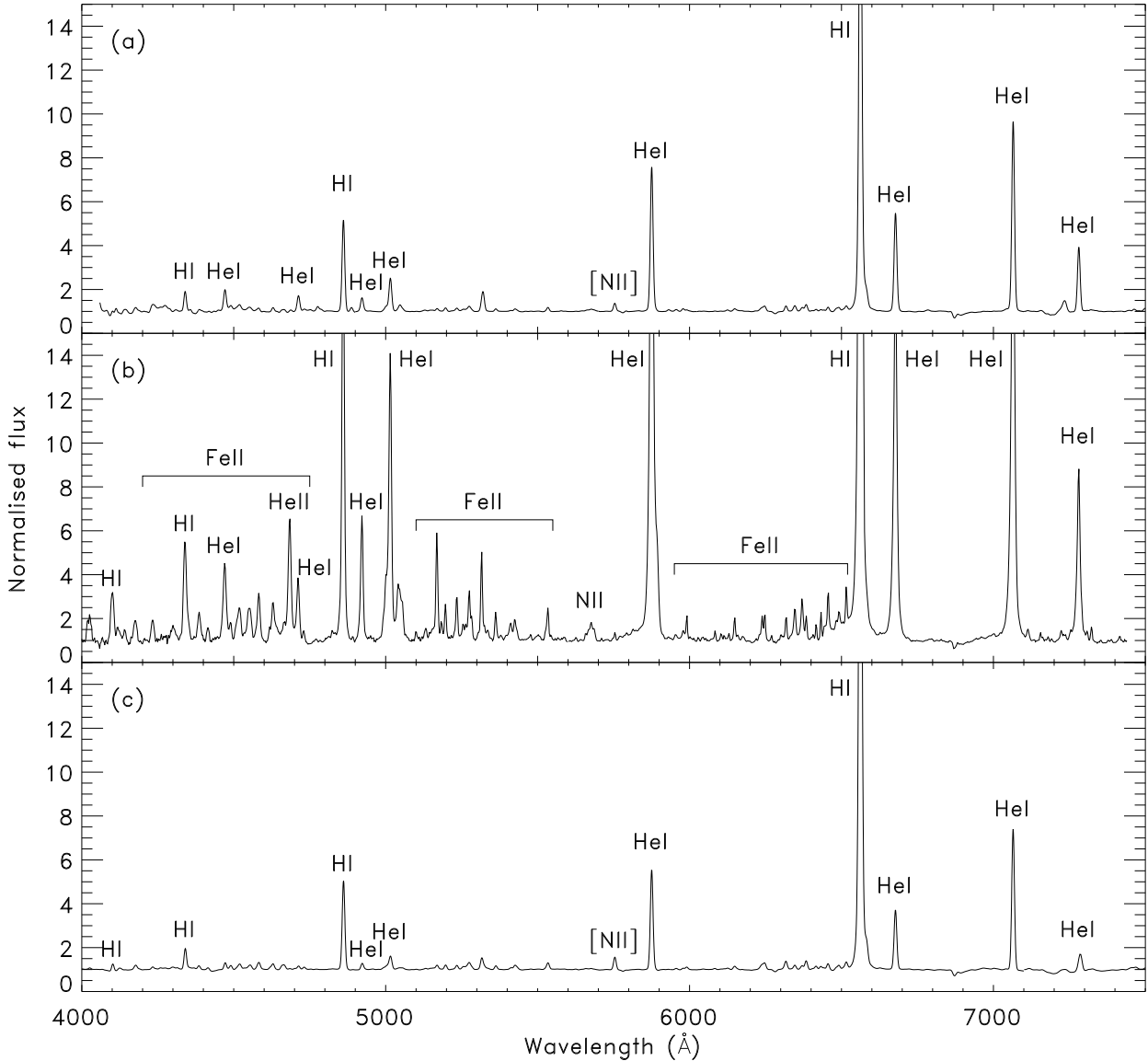}}
\caption{Comparison between normalised spectra from a) before the
outburst (1998 January 27--28), b) early in the outburst (1998 April
4), and c) after the outburst (2000 February 6 for $\lambda >
7100$\,\AA\ and 2000 December 1 elsewhere).  Major features are
labelled.  The strongest groups of Fe\,\textsc{ii} lines are also
indicated, but other species are present in these complexes and
Fe\,\textsc{ii} lines are also found elsewhere.}
\label{MasterSpecFig}
\end{figure*}

We plot the pre-outburst optical spectrum (from 1998 OHP data) in
Fig.~\ref{MasterSpecFig}a and in Table \ref{QuiescentEWTable} we
compile the equivalent widths of the strongest emission lines in
quiescence.  Coverage is clearly patchy, but no line exhibits dramatic
variations in equivalent width.  In particular the relative strengths
of Balmer and He lines remain roughly the same.  Downes
(\cite{Downes:1984a}) does not include equivalent widths, but the
relative line strengths plotted in his spectra (e.g.\ H$\gamma$ to
He\,\textsc{i} 4471) look very similar to the results in Table
\ref{QuiescentEWTable}, and also to Fig.\ \ref{MasterSpecFig}a.  It
therefore appears that the pre-outburst spectrum was stable for at
least $\sim30$ years before the outburst, although large gaps in
coverage are obviously present.  As discussed in Sect.~\ref{SpecEvol},
the post-outburst spectrum is clearly different to this, but it
remains to be seen whether the differences represent a delayed return
to the pre-outburst quiescent state, or a new different quiescent
state.

\begin{table}
\begin{tabular}{lccccc} 
\hline
\noalign{\smallskip}
\multicolumn{2}{l}{Line} 
1964--7$^1$ & 1986--7$^2$ &   1987$^3$ &  1994$^3$ & 1998$^4$ \\
\hline
\noalign{\smallskip}
H$\delta$ 4100      &     4.9 &    --   &   --   &   5.1   &--\\
H$\gamma$ 4340      &    13.4 &    10.5 &   --   &  10.9   &10.5\\
H$\beta$  4861      &    53   &    65   &   57   &  44     &46.5\\
H$\alpha$ 6562      &    --   &   241   &   --   & (397)   &$>300$\\
\noalign{\smallskip}
He\,\textsc{i}   4026       &     4.0 &    --   &   --   &   3.6   &--\\ 
He\,\textsc{i}   4471       &    10.6 &    --   &   --   &  10.7   &9.6\\
He\,\textsc{i}   4713       &     9.3 &    --   &    8.0 &   8.8   &8.0\\
He\,\textsc{i}   5875       &    93   &    --   &   --   & (106.7) &73.6\\ 
He\,\textsc{i}   6678       &    --   &    --   &   --   &  (54.5) &53.2\\
He\,\textsc{i}   7065       &    --   &    --   &   --   &  --     &89.4\\
He\,\textsc{i}   7281       &    --   &    --   &   --   &  --     &34.0\\
\noalign{\smallskip}
[N\,\textsc{ii}]  5755       &    --   &    --   &   --   &  --     & 3.5\\
\noalign{\smallskip}
\hline  
\end{tabular}
\caption{Pre-outburst equivalent widths, in \AA, of selected strong
emission lines.  Minus signs have been dropped for convenience as all
lines are in emission.  He\,\textsc{i} 4921 and 5015\,\AA\ have been
omitted as these are unresolved from strong Fe\,\textsc{ii} lines.
Sources are $^1$Chkhikvadze (\cite{Chkhikvadze:1970a}),
$^2$Miroshnichenko (\cite{Miroshnichenko:1995b}),
$^3$Section~\ref{PreoutburstSAOSection}, $^4$Jaschek \& Andrillat
(\cite{Jaschek:2000a}) and Section~\ref{OHPReductionSection}.  Lines
with no value given are not reported, outside the spectral range or
too weak to reliably measure.  Bracketed measurements were obtained
from spectra with a poorly exposed continuum rendering the equivalent
widths uncertain.}
\label{QuiescentEWTable}
\end{table}

%%%%%%%%%%%%%%%%%%%%%%%%%%%%%%%%%%%%%%%%%%%%%%%%%%%%%%%%%%%%%%%%%%%%%%%%%%%%%%%
%
\section{Outburst and post-outburst observations}
\label{OutburstSpec}
\begin{table*}
\begin{center}
\begin{tabular}{llccrcc} \hline
\noalign{\smallskip}
Date & Telescope & UT Start & UT End & Number & Wavelength & Resolution\\
     &           & &&&Range(\AA) & (\AA)\\
\noalign{\smallskip}
\hline  
\noalign{\smallskip}
{\em Pre-outburst} &&&&&&\\
05/04/87 & SAO RAS 6-m      & 23:12 & 23:50 &  1 & 4460--5005  & 1.1 \\
06/04/87 & SAO RAS 6-m      & 00:20 & 00:45 &  1 & 6620--7290  & 2.1 \\
21/01/94 & SAO RAS 6-m      & 17:38 & 18:06 &  1 & 4000--4900  & 2.2 \\
21/01/94 & SAO RAS 6-m      & 18:15 & 18:28 &  1 & 5700--6700  & 2.1 \\
26/01/98 & OHP 1.52-m       & 21:37 & 22:20 &  1 & 4060--4930  & 1.3  \\ 
27/01/98 & OHP 1.52-m       & 21:16 & 22:46 &  1 & 4860--5730  & 1.3  \\ 
27/01/98 & OHP 1.52-m       & 23:14 & 00:44 &  1 & 6250--7110  & 1.3  \\ 
28/01/98 & OHP 1.52-m       & 01:04 & 02:34 &  1 & 4060--4930  & 1.3  \\ 
28/01/98 & OHP 1.52-m       & 23:02 & 00:32 &  1 & 8040--8900  & 1.3  \\ 
29/01/98 & OHP 1.52-m       & 01:08 & 02:38 &  1 & 5560--6430  & 1.3  \\ 
29/01/98 & OHP 1.52-m       & 03:00 & 04:30 &  1 & 7060--7930  & 1.3  \\ 
\noalign{\smallskip}
{\em Outburst} &&&&&&\\
03/04/98 & FLWO 1.5-m       & 02:41 & 02:44 &  3 & 3580--7450  & 3.0  \\
03/04/98 & FLWO 1.5-m       & 02:47 & 04:19 &370 & 6040--7035  & 1.1  \\
04/04/98 & SAO RAS 6-m      & 17:07 & 17:11 &  2 & 3700--6130  & 4.0  \\
04/04/98 & SAO RAS 6-m      & 17:14 & 17:15 &  1 & 4990--7460  & 4.0  \\
05/04/98 & SAO RAS 6-m      & 15:33 & 15:35 &  1 & 3700--6130  & 4.0  \\
05/04/98 & SAO RAS 6-m      & 15:40 & 15:41 &  1 & 4990--7460  & 4.0  \\
06/04/98 & SAO RAS 6-m      & 15:47 & 16:20 &  2 & 3700--6130  & 4.0  \\
06/04/98 & SAO RAS 6-m      & 15:52 & 16:04 &  3 & 4990--7460  & 4.0  \\
09/04/98 & WHT 4.2-m        & 21:15 & 21:46 &  2 & 3870--6110  & 0.1 \\
10/04/98 & WHT 4.2-m        & 20:50 & 21:48 &  3 & 3870--6110  & 0.1 \\
11/04/98 & WHT 4.2-m        & 20:49 & 21:51 &  5 & 3870--6110  & 0.1 \\
18/04/98 & McDonald 2.7-m   & 02:47 & 05:00 & 61 & 6190--6900  & 1.2 \\
18/04/98 & FLWO 1.5-m       & 03:08 & 03:28 & 11 & 3630--7490  & 3.0  \\
19/04/98 & McDonald 2.7-m   & 02:26 & 04:07 & 65 & 6190--6900  & 1.2 \\
19/04/98 & FLWO 1.5-m       & 02:54 & 03:10 &  2 & 3740--4740  & 1.1  \\
19/04/98 & FLWO 1.5-m       & 03:08 & 03:18 &  3 & 4310--5310  & 1.1  \\
19/04/98 & FLWO 1.5-m       & 03:27 & 03:32 &  4 & 6040--7030  & 1.1  \\
20/04/98 & McDonald 2.7-m   & 02:12 & 04:54 & 80 & 6190--6900  & 1.2 \\
19/04/98 & SAO RAS 6-m      & 14:58 & 15:15 & 25 & 3700--6130  & 8.0  \\
19/04/98 & SAO RAS 6-m      & 15:19 & 15:27 &  2 & 4990--7460  & 8.0  \\
\noalign{\smallskip}
\hline
\end{tabular}
\caption{Summary of spectroscopic observations, both pre-, during, and
post-outburst which were quantitatively analysed for this work.  In
some cases multiple spectra from the same night covering the same
wavelength range have been merged on one line; in this case start and
end times refer to the first and last spectrum and do not indicate
total exposure time.  Continued on the next page.}
\label{SpecLog}
\end{center}
\end{table*}

\begin{table*}
\begin{center}
\begin{tabular}{llccrcc} \hline
\noalign{\smallskip}
Date & Telescope & UT Start & UT End & Number & Wavelength & Resolution\\
     &           & &&&Range(\AA) & (\AA)\\
\noalign{\smallskip}
\hline  
\noalign{\smallskip}
{\em Post-outburst} &&&&&&\\
16/05/98 & SAO RAS 6-m      & 18:08 & 18:21 &  3 & 3700--6130  & 4.0  \\
16/05/98 & SAO RAS 6-m      & 18:29 & 18:40 &  2 & 4990--7460  & 4.0  \\
20/07/98 & WHT 4.2-m        & 05:43 & 06:01 &  4 & 3500--7000  & 3.1  \\
21/07/98 & WHT 4.2-m        & 05:57 & 06:06 &  2 & 4800--5200  & 0.4  \\
15/09/98 & FLWO 1.5-m       & 09:49 & 09:53 &  3 & 3660--7530  & 3.0  \\
17/09/98 & FLWO 1.5-m       & 10:46 & 10:50 &  2 & 3660--7530  & 3.0  \\
18/09/98 & FLWO 1.5-m       & 11:43 & 11:47 &  2 & 3660--7530  & 3.0  \\
19/09/98 & FLWO 1.5-m       & 11:42 & 11:44 &  3 & 3660--7530  & 3.0  \\
21/09/98 & FLWO 1.5-m       & 08:54 & 08:56 &  3 & 3660--7530  & 3.0  \\
23/09/98 & FLWO 1.5-m       & 11:42 & 11:46 &  2 & 3660--7530  & 3.0  \\
24/09/98 & FLWO 1.5-m       & 11:41 & 11:45 &  2 & 3660--7530  & 3.0  \\
29/09/98 & FLWO 1.5-m       & 11:57 & 11:58 &  3 & 3630--7490  & 3.0  \\
30/09/98 & FLWO 1.5-m       & 08:37 & 08:39 &  3 & 3660--7530  & 3.0  \\
14/10/98 & FLWO 1.5-m       & 10:19 & 10:22 &  2 & 3650--7520  & 3.0 \\
15/10/98 & FLWO 1.5-m       & 10:55 & 10:58 &  2 & 3650--7520  & 3.0 \\
16/10/98 & FLWO 1.5-m       & 10:44 & 10:47 &  2 & 3650--7520  & 3.0 \\
23/10/98 & FLWO 1.5-m       & 10:49 & 10:51 &  3 & 3580--7450  & 3.0 \\
29/10/98 & FLWO 1.5-m       & 09:03 & 09:07 &  2 & 3650--7520  & 3.0 \\
01/11/98 & Asiago 1.82-m    & 02:20 & 02:35 &  1 & 6360--8060  & 3.3 \\
12/11/98 & FLWO 1.5-m       & 10:02 & 10:13 &  3 & 3630--7500  & 3.0 \\
13/11/98 & FLWO 1.5-m       & 11:41 & 11:44 &  3 & 3630--7500  & 3.0 \\
14/11/98 & FLWO 1.5-m       & 10:02 & 10:06 &  3 & 3630--7500  & 3.0 \\
15/11/98 & FLWO 1.5-m       & 08:14 & 08:17 &  3 & 3630--7500  & 3.0 \\
16/11/98 & FLWO 1.5-m       & 10:41 & 10:44 &  3 & 3630--7500  & 3.0 \\
17/11/98 & FLWO 1.5-m       & 11:08 & 11:10 &  4 & 3630--7500  & 3.0 \\
18/11/98 & FLWO 1.5-m       & 09:06 & 09:18 &  4 & 3630--7500  & 3.0 \\
19/11/98 & FLWO 1.5-m       & 09:40 & 09:44 &  3 & 3660--7530  & 3.0 \\
21/11/98 & FLWO 1.5-m       & 09:31 & 09:34 &  3 & 3660--7530  & 3.0 \\
22/11/98 & FLWO 1.5-m       & 08:35 & 08:39 &  3 & 3660--7530  & 3.0 \\
23/11/98 & FLWO 1.5-m       & 09:04 & 09:07 &  3 & 3660--7530  & 3.0 \\
24/11/98 & FLWO 1.5-m       & 10:24 & 10:27 &  3 & 3660--7530  & 3.0 \\
25/11/98 & FLWO 1.5-m       & 07:56 & 08:01 &  3 & 3660--7530  & 3.0 \\
26/11/98 & FLWO 1.5-m       & 07:47 & 07:51 &  3 & 3660--7530  & 3.0 \\
27/11/98 & FLWO 1.5-m       & 08:23 & 08:27 &  3 & 3660--7530  & 3.0 \\
30/11/98 & FLWO 1.5-m       & 11:17 & 11:22 &  3 & 3660--7530  & 3.0 \\
\hline
\end{tabular}
\addtocounter{table}{-1}
\caption{(continued)}
\label{SpecLog2}
\end{center}
\end{table*}

\begin{table*}
\begin{center}
\begin{tabular}{llccrcc} \hline
\noalign{\smallskip}
Date & Telescope & UT Start & UT End & Number & Wavelength & Resolution\\
     &           & &&&Range(\AA) & (\AA)\\
\noalign{\smallskip}
\hline  
\noalign{\smallskip}
10/12/98 & FLWO 1.5-m       & 05:59 & 06:03 &  3 & 3660--7530  & 3.0 \\
12/12/98 & FLWO 1.5-m       & 07:29 & 07:32 &  3 & 3660--7530  & 3.0 \\
13/12/98 & FLWO 1.5-m       & 05:29 & 05:33 &  3 & 3660--7530  & 3.0 \\
14/12/98 & FLWO 1.5-m       & 09:25 & 09:29 &  3 & 3660--7530  & 3.0 \\
19/12/98 & FLWO 1.5-m       & 06:07 & 06:11 &  3 & 3660--7530  & 3.0 \\
20/12/98 & FLWO 1.5-m       & 06:22 & 06:27 &  3 & 3660--7530  & 3.0 \\
21/12/98 & FLWO 1.5-m       & 07:35 & 07:39 &  3 & 3660--7530  & 3.0 \\
22/12/98 & FLWO 1.5-m       & 06:08 & 06:12 &  3 & 3660--7530  & 3.0 \\
23/12/98 & FLWO 1.5-m       & 06:11 & 06:15 &  3 & 3660--7530  & 3.0 \\
25/12/98 & FLWO 1.5-m       & 08:32 & 08:36 &  3 & 3660--7530  & 3.0 \\
26/12/98 & FLWO 1.5-m       & 04:52 & 04:56 &  3 & 3660--7530  & 3.0 \\
27/12/98 & FLWO 1.5-m       & 05:14 & 05:17 &  3 & 3660--7530  & 3.0 \\
28/12/98 & FLWO 1.5-m       & 03:14 & 03:18 &  3 & 3660--7530  & 3.0 \\
03/01/99 & OHP 1.52-m       & 00:26 & 01:55 &  1 & 8040--8900  & 1.3 \\ 
03/01/99 & OHP 1.52-m       & 03:30 & 05:30 &  1 & 4060--4930  & 1.3 \\ 
03/01/99 & OHP 1.52-m       & 20:05 & 21:35 &  1 & 4860--5730  & 1.3 \\ 
03/01/99 & OHP 1.52-m       & 22:11 & 23:41 &  1 & 5560--6430  & 1.3 \\ 
04/01/99 & OHP 1.52-m       & 00:04 & 01:34 &  1 & 6250--7110  & 1.3 \\ 
04/01/99 & OHP 1.52-m       & 02:33 & 04:33 &  1 & 7060--7930  & 1.3 \\ 
05/01/99 & Loiano 1.52-m    & 21:17 & 21:27 &  1 & 6360--8220  & 3.3 \\
29/01/00 & Calar Alto 3.5-m & 23:44 & 23:49 &  1 & 3700--6800  & 6.1 \\
06/02/00 & Loiano 1.52-m    & 21:11 & 21:41 &  1 & 3500--5400  & 5.5 \\
06/02/00 & Loiano 1.52-m    & 20:33 & 21:06 &  2 & 3530--8830  & 8.3 \\
19/07/00 & Skinakas 1.3-m   & 02:30 & 02:35 &  1 & 5550--7550  & 4.5 \\
16/10/00 & Skinakas 1.3-m   & 01:37 & 01:40 &  1 & 5250--7300  & 4.5 \\
01/12/00 & WHT 4.2-m        & 20:22 & 20:37 &  3 & 3700--7200  & 3.6 \\
01/12/00 & WHT 4.2-m        & 21:06 & 21:37 &  3 & 3870--4310  & 0.3 \\
28/04/01 & WHT 4.2-m        & 20:53 & 21:03 &  1 & 3600--4030  & 0.3 \\
28/04/01 & WHT 4.2-m        & 20:45 & 21:03 & 15 & 6300--6700  & 0.6 \\
08/08/01 & Skinakas 1.3-m   & 01:42 & 01:47 &  1 & 5460--7430  & 4.5 \\
13/09/01 & Skinakas 1.3-m   & 00:45 & 00:48 &  1 & 5230--7200  & 4.5 \\
23/10/01 & Bok 2.3-m        & 10:30 & 10:46 &  3 & 3885--5030  & 1.8 \\
\hline
\end{tabular}
\addtocounter{table}{-1}
\caption{(continued)}
\label{SpecLog3}
\end{center}
\end{table*}

We next present a series of spectroscopic observations obtained during
1998 April and May, covering the outburst and immediate
aftermath. Spectroscopy taken during this time indicates the presence
of high excitation lines of He\,\textsc{ii} at the time of the
outburst which fade on a timescale $\sim10$\,days; similar variability
is seen in most other lines observed during this period. Such
behaviour occurs over longer timescales than the X-ray outburst; {\it
XTE} observations reveal that the X-ray flux had already dropped by a
factor of a hundred by April 4 (Belloni et al.\ \cite{Belloni:1999a}).

We obtained both blue and red spectra during the outburst, thus
enabling us to follow the evolution of the outburst and determine
changes in the circumstellar environment during this period.  Spectra
were obtained from four telescopes in the month following the X-ray
outburst: the 1.5\,m at the FLWO, the 6\,m telescope of the SAO RAS,
the 4.2\,m WHT on La Palma and the McDonald Observatory 2.7\,m.
Subsequent observations obtained during the period between 1998 May
and 2001 October constrain the long term post-outburst behaviour of
the system. All of these observations are summarised in Table
\ref{SpecLog} and more details are given in the following subsections.

\subsection{F. L. Whipple Observatory observations}
A series of spectra were obtained using the 1.5\,m telescope at the
F. L. Whipple Observatory (FLWO) and the FAST spectrometer.  These
observations began soon after the outburst was discovered and
continued until the end of 1998.  The spectra were obtained with two
different gratings, of 300 and 1200\,lines\,mm$^{-1}$, giving 3.0 and
$1.1$\AA\ resolution respectively.

These observations included a series of $370$ 2\,s spectra with $\sim
1.1$\AA\ resolution and a dispersion of 0.4\,\AA\,pix$^{-1}$, covering
6040--7035\,\AA\ which were obtained on the night of 1998 April 2.  We
searched for shifts in the velocities of the emission lines by
cross-correlating all the spectra against a single template.  The
largest shifts are $\pm 12$\,km\,s$^{-1}$, the rms in the velocities
is 3.6\,km\,s$^{-1}$.  Given the high airmass (from 1.5 to 2.0) and
the $2.0$\,arcsec ~slit, these variations are likely due to variable
illumination of the slit rather than intrinsic motions in the emission
lines, as the velocity resolution was about 50\,km\,s$^{-1}$.
Nonetheless, we searched for periodicities in the velocities using the
algorithm of Stellingwerf (\cite{Stellingwerf:1978a}), but found
nothing significant.  We will further discuss evidence for variations
in line strengths and profiles in Sect.~\ref{Rapid}.
\subsection{Special Astrophysical Observatory observations}
Outburst observations spanning 1998 April to May were made using the
6\,m BTA telescope of the SAO RAS.  A medium resolution spectrograph,
SP--124 (at the Nasmyth--1 focus), was used with a Photometrix
1024$\times$1024 CCD detector and the B1 grating
(600\,lines\,mm$^{-1}$ ) providing a spectral dispersion of
2.4\,\AA\,pixel$^{-1}$ in two overlapping spectral ranges 3700--6100
and 5000--7500\,\AA. The slit width employed was either 1\,arcsec or
2\,arcsec, depending on seeing conditions, providing a spectral
resolution of $\sim4$\,\AA\ or $\sim8$\,\AA\ respectively.  Primary
reduction of the CCD spectra including corrections for the dark
current, debiasing and sky subtraction was accomplished with {\sc
midas} (1996 Nov.\ version).  He--Ne--Ar lamps were used for
wavelength calibration.
\subsection{WHT/UES observations}
\object{CI~Cam} was observed at high resolution using the Utrecht
Echelle Spectrograph (UES) on the William Herschel Telescope (WHT) on
La Palma on 1998 April 9--11.  Spectra were extracted using a simple
(non-optimal) extraction with sky subtraction using {\sc echomop}
software.  Cosmic rays were removed by hand.
\subsection{McDonald Observatory observations}
Intermediate dispersion spectroscopy was obtained on the nights of
1998 April 18--20 using the Large Cassegrain Spectrometer (LCS) on the
McDonald Observatory 2.7\,m telescope.  Each night involved a series
of observations spanning about 2\,hours to search for variability;
this will be discussed further in Sect.~\ref{Rapid}.  Reduction and
extraction were performed using standard {\tt iraf} tasks.
Observations of neon arc lines were performed at 20\,min intervals to
provide wavelength calibration.  Corrections were made to the
interpolated wavelength solution using night sky emission lines.  As
for the FLWO observations, no significant wavelength variations were
found.
\subsection{WHT/ISIS observations}
Intermediate and high dispersion spectroscopy was obtained on several
nights from 1998--2001 using the ISIS dual-beam spectrograph on the
WHT.  Most exposures used the blue arm with the EEV10 or EEV12 CCD.
The second spectrum from 2001 April 28 used the red arm with the TEK4
CCD.  Slit widths were 0.7--1.0\,arcsec.  Reduction and extraction
were performed using standard {\tt iraf} tasks.  Copper-neon and/or
copper-argon lamps were observed to provide wavelength calibration.
No flux calibration of these spectra was attempted.
\subsection{Loiano \& Asiago Observatory observations}
Post-outburst spectra of the source were obtained in 1999 January 5
and 2000 February 6 using the 1.52\,m G. D. Cassini telescope at the
Loiano Observatory (Italy). The telescope was equipped with the
Bologna Faint Object Spectrograph and Camera (BFOSC) and a
2048$\times$2048 Loral CCD. In January 1999 grism \#8 was used, giving
a resolution of $\sim 3$\AA, and in February 2000 the low-dispersion
grisms \#3 (blue) and \#4 (red) were used. One further spectrum was
taken on 1998 November 1st using the 1.82\,m telescope at Asiago
Observatory (Italy), equipped with the Asiago Faint Object
Spectrograph and Camera (AFOSC) and a 1024$\times$1024 thinned SITe
CCD. This instrument is identical to BFOSC and grism \#8 was used,
giving the same resolution.
\subsection{Haute Provence observations}
Post-outburst spectra obtained from OHP in 1999 used the same
configuration as the 1998 pre-outburst observations, hence the details
are the same as described in Sect.~\ref{OHPReductionSection}.
\subsection{Calar Alto observations}
One low-resolution flux-calibrated spectrum was taken on 2000 January
29 using the blue arm of the TWIN spectrograph on the 3.5\,m telescope
at Calar Alto (Spain) and a 2000$\times$800 thinned SITe CCD.
\subsection{Skinakas Observatory observations}
Spectra taken on 2001 July 19 \& October 16 were taken with the 1.3\,m
f/7.7 Ritchey-Chretien telescope at Skinakas Observatory (Crete,
Greece).  This was equipped with a 2000$\times$800 ISA SITe CCD and a
1302\,lines\,mm$^{-1}$ grating (using a 80$\mu$m slit), giving a
dispersion of $\sim$1\,\AA\,pixel$^{-1}$.
\subsection{Steward Observatory observations}
Finally three spectra were obtained using the Bok 2.3\,m telescope of
the Steward Observatory with the Boller and Chivens long slit
spectrograph and a 1200$\times$800 CCD.  A first order
1200\,lines\,mm$^{-1}$ grating was used and there were no filters.
Flux calibration was not reliable as there were significant slit
losses.
%
%%%%%%%%%%%%%%%%%%%%%%%%%%%%%%%%%%%%%%%%%%%%%%%%%%%%%%%%%%%%%%%%%%%%%%%%%%%%%%%
%
\section{Interstellar extinction and distance estimates}
\label{ExtinctionSection}
There remain large uncertainties about the interstellar extinction to
\object{CI~Cam}, and even more so concerning the distance.  A number
of extinction estimates have been made but these are complicated by
the uncertainty about how much of the measured extinction is intrinsic
to the source; different methods may be more or less sensitive to this
depending on what is actually measured.  Consequently it is of value
to compare as many independent methods as possible.  We discuss only
recent estimates as these are likely to use higher quality data and/or
more sophisticated models, superseding earlier estimates (e.g.\
Chkhikvadze \cite{Chkhikvadze:1970a}).

A number of attempts have been made to fit the broad band optical--far
IR spectral energy distribution (SED) using a model including
extinction as a free parameter.  Belloni et al.\
(\cite{Belloni:1999a}) fitted the SED using a Kurucz atmosphere and a
dust model and derived $E(B-V)=1.18\pm0.04$\,mag and hence
$A_V\sim4.4$\,mag.  Clark et al.\ (\cite{Clark:2000a}) used a similar
model to derive $A_V\sim3.71$\,mag.  Zorec (\cite{Zorec:1998a}) used a
model in which the interstellar extinction and distance were
constrained to be consistent rather than independent, and estimated
$d\sim1.75$\,kpc, $E(B-V)_{\rm ism}\sim0.8$\,mag and $A_{V, \rm
local}\sim 2.4$\,mag, implying a total extinction somewhat higher than
the other two estimates.

Orlandini et al.\ (\cite{Orlandini:2000a}) used emission line ratios
to estimate $E(B-V)=1.54$\,mag (from He lines) or $E(B-V)=1.02$\,mag
(from H lines).  This is based on theoretical predictions of the
ratios of the strongest lines in \object{CI~Cam}.  In Be stars and
other early OB type stars with dense winds, and likely also in sgB[e]
stars, these lines are subject to effects of non-local thermodynamic
equilibrium (NLTE) so this method will not be very reliable.

Robinson et al.\ (\cite{Robinson:2002a}) use the 2175\,\AA\
interstellar absorption feature in post-outburst {\it HST} data to
measure $E(B-V)=0.85\pm0.05$\,mag and $A_{V}=2.3\pm0.3$\,mag; the
large error on $A_V$ reflects the uncertainty in the choice of
extinction curve parameterised by $R_V = A_V / E(B-V)$.

\begin{figure}
\resizebox{\hsize}{!}{\rotatebox{90}{\includegraphics{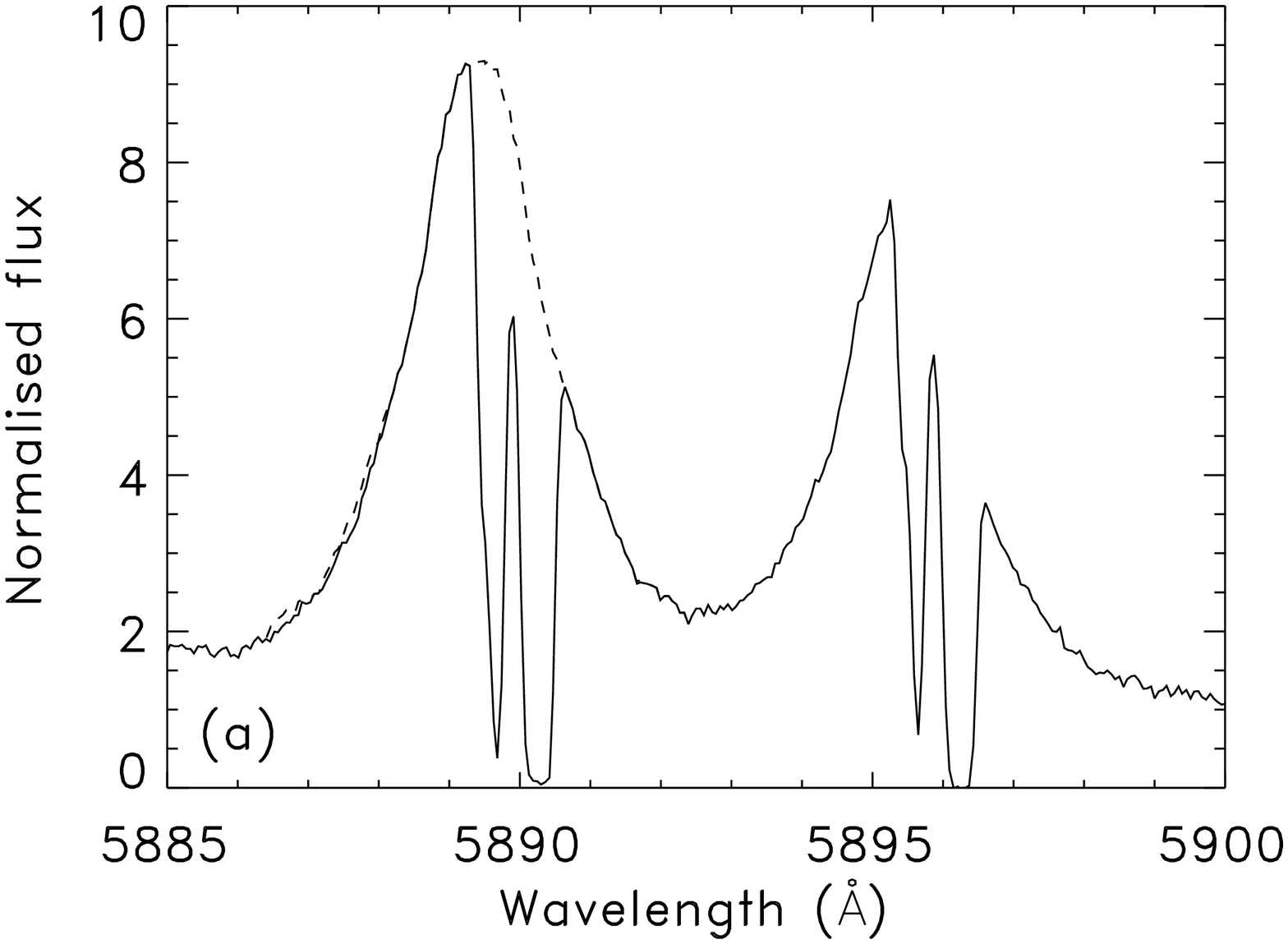}}}
\resizebox{\hsize}{!}{\rotatebox{90}{\includegraphics{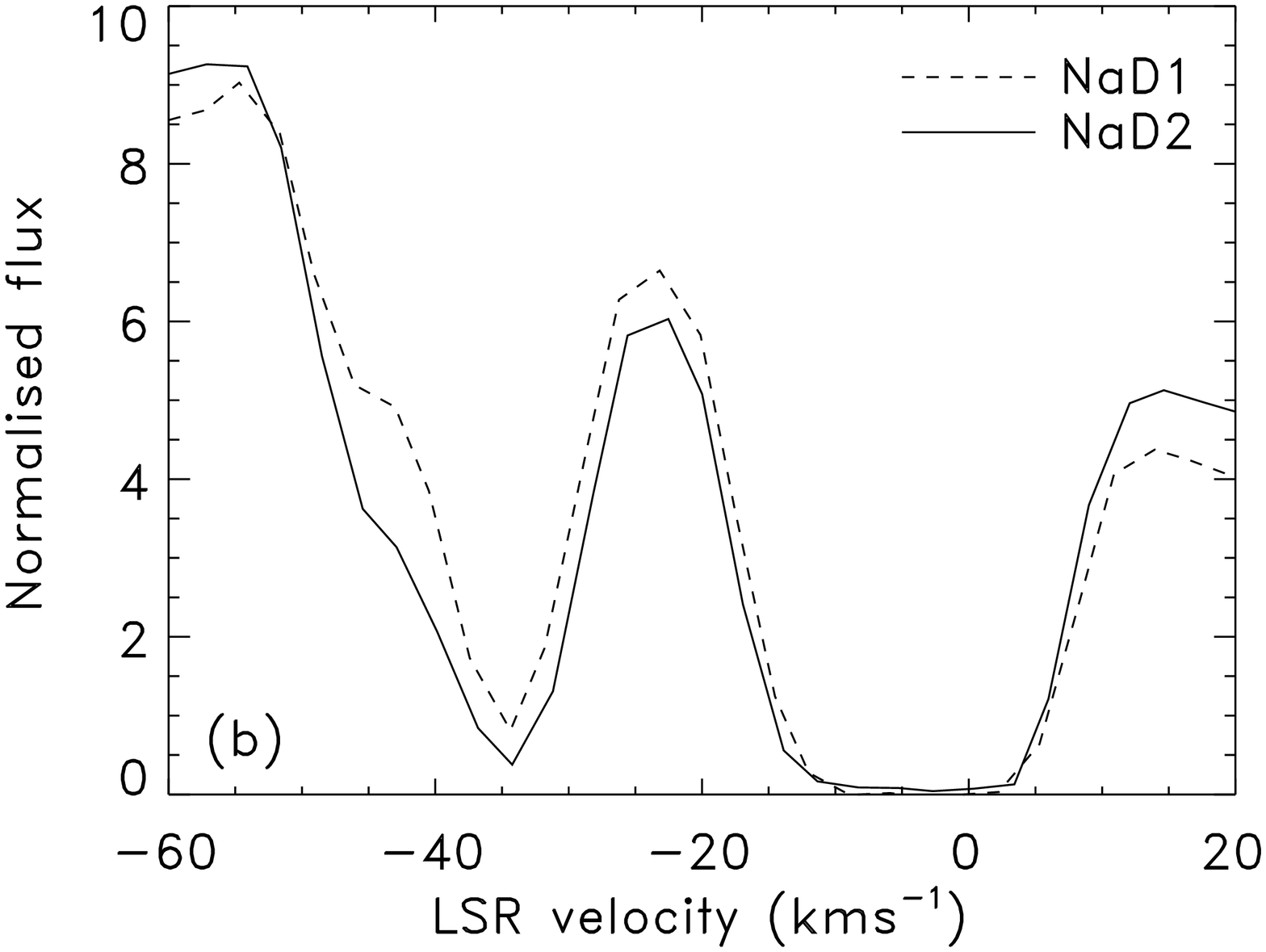}}}
\caption{a) Na\,D line profiles observed during outburst (1998 April
9).  The two interstellar absorption components are clearly visible;
the longer wavelength one is saturated.  The dashed line shows the
reconstructed Na\,D2\ line profile assuming it is symmetrical. b)
Closeup of the absorption components indicating the velocities with
respect to the LSR.  The structure in the two lines is consistent.  In
this direction the LSR velocity is expected to become more negative
with distance, so the distance increases to the left.  A quantitative
distance scale depends on the assumed rotation curve of the Galaxy.}
\label{NaDFig}
\end{figure}

We can attempt to estimate the Na\,D equivalent width, although as
noted by Munari \& Zwitter (\cite{Munari:1997a}) this is insensitive
for reddened objects as it tends to saturate.  It is also complicated
by the presence of Na\,D emission.  Fortunately, our echelle spectra
obtained in outburst resolve the Na\,D components as sharp absorptions
within the emission line (Fig.~\ref{NaDFig}a).  Reconstructing the
unabsorbed line profile is obviously uncertain, but assuming it is
symmetric we can estimate that the equivalent widths of the two Na\,D2
(5890\,\AA\footnote{We note that Munari \& Zwitter \cite{Munari:1997a}
incorrectly label the 5890\,\AA\ line as Na\,D1, and 5896\,\AA\ as
Na\,D2.  Their work is, however, internally consistent; Munari, priv.\
comm.})  components observed are 0.36 and 0.52\,\AA.  The latter
component does appear saturated, so this only provides a lower-limit
to the reddening.  If the profile is actually asymmetric with the same
extended blue wing seen in many other lines then this will increase
the inferred equivalent width, implying a reddening further above our
lower limit.  Using the calibration of Munari \& Zwitter
(\cite{Munari:1997a}) for each of the two components we derive a lower
limit for the combined reddening of $E(B-V)\ga0.5$\,mag.

Our spectra also reveal several diffuse interstellar bands (DIBs).
Our principal reddening calibrator is the $\lambda$5780 band, using
the calibration of Herbig (\cite{Herbig:1993a}).  Using the WHT/UES
spectra we measure an EW of $320\pm30$\,m\AA\ implying
$E(B-V)=0.64\pm0.06$\,mag.  The DIB at 5797\,\AA\ is too blended with
emission lines to reliably measure but we can approximately estimate
$140\pm30$\,m\AA\ implying $E(B-V)=0.9\pm0.2$\,mag (Herbig
\cite{Herbig:1993a}).  For 5705\,\AA\ we measure $110\pm20$\,m\AA\
implying $E(B-V)=0.49\pm0.09$\,mag (Herbig \cite{Herbig:1975a}).  We
also use an average of several WHT/ISIS and Skinakas spectra to
estimate an equivalent width of $180\pm30$\,m\AA\ for $\lambda6203$
implying $E(B-V)=0.62\pm0.11$\,mag (Herbig \cite{Herbig:1975a}), and a
central depth of 0.90--0.97 for the very broad $\lambda4428$ feature
implying $0.3<E(B-V)<1.1$ (Krelowski et al.\ \cite{Krelowski:1987a}).
The 6284\,\AA\ line may also be present but is complicated by Telluric
absorption.  All of the reliable DIB measurements are consistent with
$E(B-V)\sim0.65$\,mag as suggested by our principal calibrator,
5780\,\AA.

To summarise, methods based on interstellar absorption features
(2175\,\AA, Na\,D, DIBs) all suggest $E(B-V)\sim0.5-1.0$\,mag, and
hence visual extinction $A_V\sim1.5-3$\,mag.  Methods based on fitting
the SED or line ratios, however, favour larger total extinction
$A_V\sim4$\,mag.  This could arise if the absorption features are only
produced in the interstellar medium, and not by local dust extinction,
as a consequence of different chemical compositions and/or grain
sizes.  The SED methods are all model dependent, however, and it may
be that the models used are either incorrect or incomplete.  This
leaves us with some ambiguity about how to deredden spectra of
\object{CI~Cam}.  The higher values inferred from SED modelling are
most relevant, in the sense that they directly measure the distortion
of the broad band spectrum, but they are also model dependent.
Measurements based on absorption features are more objective, but do
not measure the same thing.

Distance estimates for \object{CI~Cam} are plagued by even more
uncertainty than extinction estimates.  A number of works have derived
distances of 1--2\,kpc (Chkhikvadze \cite{Chkhikvadze:1970a}; Zorec
\cite{Zorec:1998a}; Belloni et al.\ \cite{Belloni:1999a}; Clark et
al.\ \cite{Clark:2000a}).  All of these, however, involve uncertain
assumptions about the luminosity class of the star, or the relation
between interstellar absorption and distance.  Robinson et al.\
(\cite{Robinson:2002a}) challenged this conclusion, arguing for a
larger distance.  Based on spectroscopic similarities to the largest
and most luminous sgB[e] stars they concluded that the distance had to
be larger than 2\,kpc.  Since the line of sight passes well above the
warped Galactic plane for 2--6\,kpc, they argued that a young object
like \object{CI~Cam} must lie beyond rather than within this distance
range.  In support of this they note that its radial velocity is
consistent with a distance of $\sim7$\,kpc, assuming differential
Galactic rotation.  These arguments, however, make the assumptions
that \object{CI~Cam} is comparable to the most luminous sgB[e] stars,
and that it does not have significant peculiar velocity.

More objective constraints on the distance are harder to obtain.  We
can, however, make one estimate which is almost completely independent
of what \object{CI~Cam} is.  As already described, the Na\,D lines
show sharp absorption components.  The velocities of these in the
local standard of rest (LSR) are $-4.5\pm0.6$\,km\,s$^{-1}$ and
$-34.8\pm0.4$\,km\,s$^{-1}$ for the stronger and weaker components
respectively (see Fig.~\ref{NaDFig}b).  The stronger component is
saturated so it could itself involve multiple components with LSR
velocities up to $-10$\,km\,s$^{-1}$.  The weaker component may also
have an additional component in its blue wing at around
$-45$\,km\,s$^{-1}$; both the Na\,D1 and Na\,D2 lines suggest such a
feature at this velocity.  All of these components are somewhat
redshifted with respect to \object{CI~Cam}, which has an LSR velocity
of $-51$\,km\,s$^{-1}$ (Robinson et al.\ \cite{Robinson:2002a}).  It
is therefore unlikely they are associated with circumstellar material;
much more likely is that they are interstellar gas.  If so then we
expect them to be moving in the plane in near circular orbits
following Galactic rotation, and their velocities can be used to
estimate their distances, and hence a lower limit on the distance of
\object{CI~Cam}.  There are obviously uncertainties introduced by
non-circular motions and an imperfectly known rotation curve.  We
estimate distances using the range of rotation curves illustrated by
Olling \& Merrifield (\cite{Olling:1998a}), with the spread in values
giving an estimate of the uncertainty in the measurement.  The
saturated component is clearly associated with the Local Arm, with a
maximum velocity $-10$\,km\,s$^{-1}$ corresponding to a maximum
distance $\sim1$\,kpc.  We might expect the next feature to correspond
to the Perseus Arm at a distance $\sim2.5$\,kpc, but the Perseus Arm
is not well defined in the direction of \object{CI~Cam} as our line of
sight passes well above the Galactic plane at that point (c.f.\
discussion by Robinson et al.\ \cite{Robinson:2002a}), so it is
unsurprising that it produces no absorption feature.  The next feature
out, at $-35$\,km\,s$^{-1}$, is at an implied distance of
3.9--5.6\,kpc, and probably corresponds to the next spiral arm out,
where our line of sight passes back into the warped outer disc;
\object{Cam~OB3} (Humphreys et al.\ \cite{Humphreys:1978a}) likely
belongs to this arm.  If the third weak feature, at
$-45$\,km\,s$^{-1}$, is real and also indicative of Galactic rotation,
then it suggests an even greater minimum distance of 6--8\,kpc,
dependent on the assumed rotation curve.

An alternative empirical approach is to use the radial velocity maps
of Brand \& Blitz (\cite{Brand:1993a}), derived from H\,\textsc{ii}
regions and reflection nebulae.  This will take account of
non-circular motions.  If anything, these suggest an even larger
distance.  The coverage in the direction of \object{CI~Cam} is sparse,
but representative distant objects from their sample (S208, S211 and
S212) have LSR velocities between -30 and -38\,km\,s$^{-1}$ and
distance estimates of 5.9--7.6\,kpc.

The distance implied for \object{CI~Cam} by the interstellar features
is thus large; it is at least 4\,kpc, and may well be beyond 6\,kpc.
This conclusion is essentially consistent with that derived
independently by Robinson et al.\ (\cite{Robinson:2002a}) using
different methods.

If the closest distance is adopted it is possible that \object{CI~Cam}
could be associated with \object{Cam~OB3}, possibly as a runaway
object.  It lies $\sim2\fdg7$ from the centre of the association; the
members identified by Humphreys (\cite{Humphreys:1978a}) lie at up to
$1\fdg2$ from the centre.  The distance modulus adopted by Humphreys
(\cite{Humphreys:1978a}) was 12.6 (3.3\,kpc).  \object{CI~Cam}
therefore lies somewhat outside the association, and appears to be
further away, but we cannot rule out the possibility that it is a
runaway from \object{Cam~OB3}.  Alternatively it could lie beyond the
spiral arm containing \object{Cam~OB3}.

Where it is necessary to assume a value for the reddening in what
follows, we either take $E(B-V)=1.3$\,mag ($A_V=4$\,mag) or a range of
$0.65 < E(B-V) < 1.4$ ($2.0 < A_V < 4.4$).  For the distance, we
follow Robinson et al.\ (2002) in adopting 5\,kpc as a representative
estimate, although we agree that it could be somewhat larger than
this.  Where a lower limit is more appropriate we assume $d>4$\,kpc.
%
%%%%%%%%%%%%%%%%%%%%%%%%%%%%%%%%%%%%%%%%%%%%%%%%%%%%%%%%%%%%%%%%%%%%%%%%%%%%%%
%
\section{The nature of the components of \object{CI~Cam}}
\label{CompanionSection}

\object{CI~Cam} clearly shows the observational characteristics of the
B[e] phenomenon (Allen \& Swings \cite{Allen:1976a}; Lamers et al.\
\cite{Lamers:1998a}): strong Balmer emission lines (e.g.\
Fig.~\ref{MasterSpecFig}), low excitation permitted lines (e.g.\
Fig.~\ref{MasterSpecFig}), optical forbidden lines of
[Fe\,\textsc{ii}] and [O\,\textsc{i}] (e.g.\ Fig.~\ref{TitaniumFig}a)
and a strong infrared excess (Clark et al.\ \cite{Clark:2000a});
indeed \object{CI~Cam} has long been considered a B[e] star and was
included in the sample of Allen \& Swings \cite{Allen:1976a}.  In view
of the distance estimates discussed above \object{CI~Cam} clearly
falls within the sgB[e] sub-class. The primary characteristic of an
sgB[e] star, a luminosity of above $10^4$\,L$_{\odot}$ (Lamers et al.\
\cite{Lamers:1998a}) is satisfied for any distance above 0.8\,kpc and
for the lower limit of 4\,kpc that we have argued for above, this
rises to $10^{5.4}$\,L$_{\odot}$.  Robinson et al.\
(\cite{Robinson:2002a}) also note spectroscopic similarities to the
most luminous sgB[e] stars, consistent with this.  The other known
types of stars showing the B[e] phenomenon are the pre-main sequence
Herbig Ae/Be (HAeB[e]) stars, compact planetary nebula B[e] (cPNB[e])
stars and symbiotic B[e] (symB[e]) stars.  HAeB[e] stars typically
have luminosities of $\la10^{4.5}$\,L$_{\odot}$, are associated with
star forming regions and sometimes show evidence of infall, e.g.\
inverse P~Cygni profiles.  \object{CI~Cam} is more luminous than this
and exhibits a steep decrease in the IR flux longward of
$\sim10$\,$\mu$m (Clark et al.\ \cite{Clark:2000a}), indicative of the
absence of the cold dust that would be expected around an HAeB[e]
star.  Lamers et al.\ (\cite{Lamers:1998a}) did suggest that
\object{CI~Cam} could be a cPNB[e] star, but as these objects have
lower luminosities ($\la10^4$\,L$_{\odot}$) this is also ruled out by
the distance estimates discussed in Section~\ref{ExtinctionSection}.
SymB[e] stars are binaries also containing a cool giant which is
usually seen in the red or infrared spectrum.  Some authors have
identified \object{CI~Cam} as a symbiotic (e.g.\ Barsukova et al.\
\cite{Barsukova:2002a}), but late type features are not seen in CI~Cam
(except for the report of Miroshnichenko \cite{Miroshnichenko:1995b}
which was not corroborated by any other observations.)  We believe
that both the observed spectrum and the high luminosity therefore
identify CI~Cam as an sgB[e] star, as argued by Robinson et al.\
\cite{Robinson:2002a}.

\begin{figure}
\resizebox{\hsize}{!}{\rotatebox{90}{\includegraphics{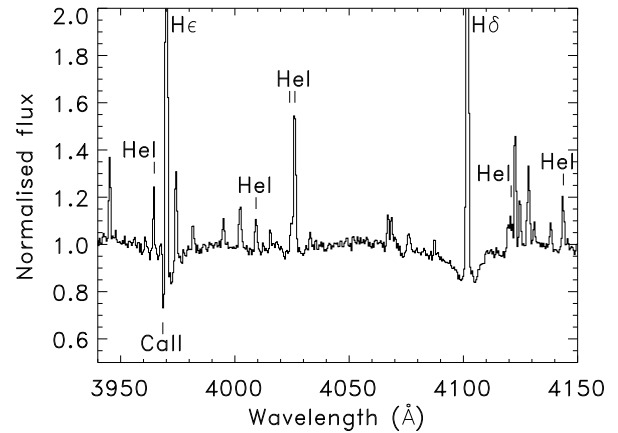}}}
\caption{Absorption wings surrounding H$\delta$ and H$\epsilon$ lines,
from 2000 December 1 WHT data.  Unmarked lines are mostly Fe\,\textsc{ii}.}
\label{BalmerAbsFig}
\end{figure}

The sgB[e] class itself includes a range of objects with likely
spectral types spanning B0--B9, and luminosities from
$10^{4.2}$--$10^{6.1}$\,L$_{\odot}$ (Lamers et al.\
\cite{Lamers:1998a}).  It is therefore of interest to attempt a more
precise classification for CI~Cam.  We might hope that the detection
of photospheric lines from the sgB[e] star would be of great help.  In
the highest resolution post-outburst spectra of CI~Cam, broad
absorption wings are seen around higher order Balmer lines, most
prominently H$\delta$ and H$\epsilon$ (Fig.~\ref{BalmerAbsFig}).
These may well be photospheric absorption lines from the sgB[e] star.
Unfortunately we cannot obtain any detailed diagnostics from these
lines because of the heavy contamination we see in the wings of the
lines (both in terms of other lines and also probably residual wind
emission); hence it is impossible to quantify the underlying
photospheric spectrum.  Indeed, the photospheric spectrum may be
poorly defined for a star with a very high mass loss rate for which a
bona fide photospheric radius is hard to determine.

While the spectral type of the star is difficult to determine from the
optical spectrum due to heavy contamination by the circumstellar
material, the detection of P~Cygni profiles in the UV resonance lines
with absorption to $\sim1000$\,km\,s$^{-1}$ (Robinson et al.\
\cite{Robinson:2002a}) is more useful.  By analogy with other sgB[e]
stars we expect that the UV P~Cygni profiles arise from a hot, polar
wind which is similar to that of normal hot supergiants.  The outflow
velocity implied for \object{CI~Cam} is directly comparable to that
seen for other early B supergiants.  The Si\,{\sc iv} 1394, 1402{\AA}
doublet functions as a powerful probe of temperature and luminosity
for B stars (Walborn, Parker, \& Nichols
\cite{Walborn:1995a}). Early-B supergiants show a strong P~Cygni wind
profile, which evolves into pure absorption for mid-B stars and is
absent in late B stars of all luminosity classes -- the P~Cygni
profiles therefore suggest an early B classification.  A similar
conclusion can be drawn from the presence of P Cygni profiles for the
C\,{\sc iv} 1549,51\,\AA\ doublet which is seen for supergiants of
type B4 or hotter (and is seen in absorption for hot B0--2 dwarfs and
giants and cooler supergiants).

By comparison of the optical emission lines with other sgB[e] stars
and luminous blue variables (c.f.\ Miroshnichenko
\cite{Miroshnichenko:1996a}), we can further narrow the spectral type
to B0--B2, as earlier spectral types show stronger He\,\textsc{ii}
emission than the very weak pre-outburst feature that we see, and
later types show He\,\textsc{i} in absorption rather than emission.

To summarise, the `normal' star is an sgB[e] star, with likely
spectral type B0--B2 and a luminosity of at least
$10^{5.4}$\,L$_{\odot}$, placing it among the hotter, more luminous
sgB[e] stars (c.f.\ Robinson et al.\ \cite{Robinson:2002a}).

The nature of the compact object implicated in the outburst is even
more uncertain than that of the normal star.  It is widely assumed
from the X-ray and $\gamma$-ray outburst observed that this must be a
black hole or neutron star (e.g.\ Belloni et al.\
\cite{Belloni:1999a}; Robinson et al.\ \cite{Robinson:2002a}).
Orlandini et al.\ (\cite{Orlandini:2000a}), however, have argued for a
thermonuclear runaway on the surface of a white dwarf.  If the 4\,kpc
lower limit on the luminosity is correct, however, then the X-ray
outburst was extremely luminous, $L_{\rm X} \ga 2 \times
10^{38}$\,erg\,s$^{-1}$.  This corresponds to the Eddington limit for
a neutron star, and for larger distances then a black hole becomes
more likely.  However, neither the X-ray spectral shape (Orr et al.\
\cite{Orr:1998a}; Ueda et al.\ \cite{Ueda:1998a}; Revnivtsev et al.\
\cite{Revnivtsev:1999a}; Belloni et al.\ \cite{Belloni:1999a}) nor the
lack of rapid X-ray variability (Frontera et al.\
\cite{Frontera:1998a}; Belloni et al.\ \cite{Belloni:1999a}) are
typical of accreting black holes or neutron stars.  Equally, while
there may be a hint of a flaring soft component (Frontera et al.\
\cite{Frontera:1998a}; Ueda et al.\ \cite{Ueda:1998a}), the X-ray
spectrum is on the whole quite hard, and not dominated by a supersoft
component as might be expected for a high luminosity accreting white
dwarf.  Consequently the identification must remain uncertain.
%
%%%%%%%%%%%%%%%%%%%%%%%%%%%%%%%%%%%%%%%%%%%%%%%%%%%%%%%%%%%%%%%%%%%%%%%%%%%%%%
%
\section{The spectral flux distribution}
\label{FluxSection}
\label{LineContributionSection}
Clark et al.\ (\cite{Clark:2000a}) considered the broad band spectral
energy distribution from photometric measurements and examined colour
changes during the outburst.  They found that the source became
systematically redder when it was brighter.  Given the large
contribution of emission lines to the spectrum, however, it is useful
to attempt to isolate continuum changes, and quantify the line
contribution.

\begin{figure}
\resizebox{\hsize}{!}{\includegraphics[angle=90]{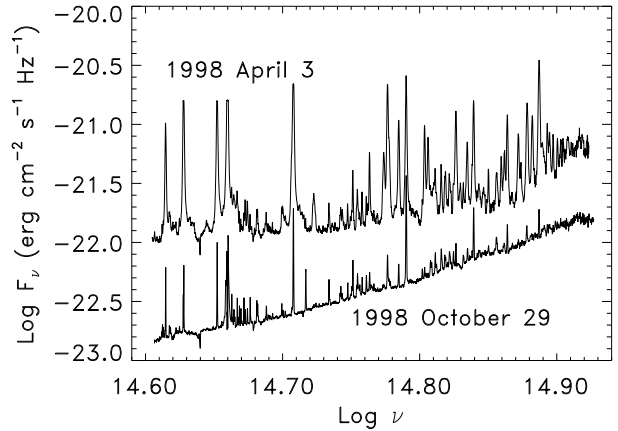}}
\caption{Flux calibrated spectra obtained from FLWO early in outburst
and after it.  These have been dereddened using the Fitzpatrick
(\cite{Fitzpatrick:1999a}) extinction curve assuming $A_V=4.0$.  In
outburst, the lines become much stronger and broader, Balmer jump
emission appears, and the continuum becomes redder.  Note that these
are long exposure spectra to ensure that the continuum is well
defined, so the stronger emission lines are saturated.}
\label{FluxSpecFig}
\end{figure}

In Fig.~\ref{FluxSpecFig} we show the first outburst spectrum
obtained, which fortunately was flux calibrated, together with a
comparable post-outburst one.  It is clear that the redder spectrum
seen in outburst was not just due to line changes, but that the
underlying continuum is redder.  We will further discuss the origin of
this enhanced continuum emission in
Section~\ref{ContinuumOriginSection}.  There is also an enhancement at
higher frequencies as the Balmer jump appears in emission, which
explains why the $(U-B)$ index remained constant during outburst
(Clark et al.\ \cite{Clark:2000a}).

In principle, we could also use flux calibrated spectra to estimate
how much of the flux in a given bandpass comes from lines.  However,
in many cases lines are saturated in these spectra, so we instead use
unsaturated, but uncalibrated spectra.  To do this we construct a flux
distribution by interpolating between the outburst JKT points (Hynes
et al.\ \cite{Hynes:1998a}) and the mean post-outburst photometry
(Clark et al.\ \cite{Clark:2000a}).  We can then multiply such a flux
distribution by a continuum normalised spectrum and perform synthetic
photometry, iterating until we have a flux distribution such that the
synthetic photometry matches the observed values.  This is very crude,
but for estimating the contribution of lines within a bandpass, the
effect of the assumed flux distribution is only to introduce a
wavelength dependent weighting within the bandpass, so the results are
weakly sensitive to the flux distribution assumed.  The contribution
from the lines can be estimated by comparing synthetic photometry
performed on the flux distribution alone with that performed on the
flux distribution multiplied by the normalised spectrum.  The
differences in magnitudes which we infer in outburst (1998 April 4)
are $\Delta B \sim 0.6$, $\Delta V \sim 0.6$ and $\Delta R \sim 0.9$,
and post-outburst $\Delta B \sim 0.08$, $\Delta V \sim 0.10$ and
$\Delta R \sim 0.35$.  While the quiescent values are fairly small,
and hence the quiescent photometry is dominated by continuum, the
outburst values can be large; the $R$ band flux, for example, is about
50\,percent line emission, due to the strong H$\alpha$ line.  These
values are similar to those quoted by Barsukova et al.\
(\cite{Barsukova:2002a}).
%
%%%%%%%%%%%%%%%%%%%%%%%%%%%%%%%%%%%%%%%%%%%%%%%%%%%%%%%%%%%%%%%%%%%%%%%%%%%%%%
%
\section{Spectral line behaviour}
\label{SpecEvol}
\begin{figure*}
\resizebox{\hsize}{!}{
\begin{tabular}{lll}
\includegraphics[width=2in]{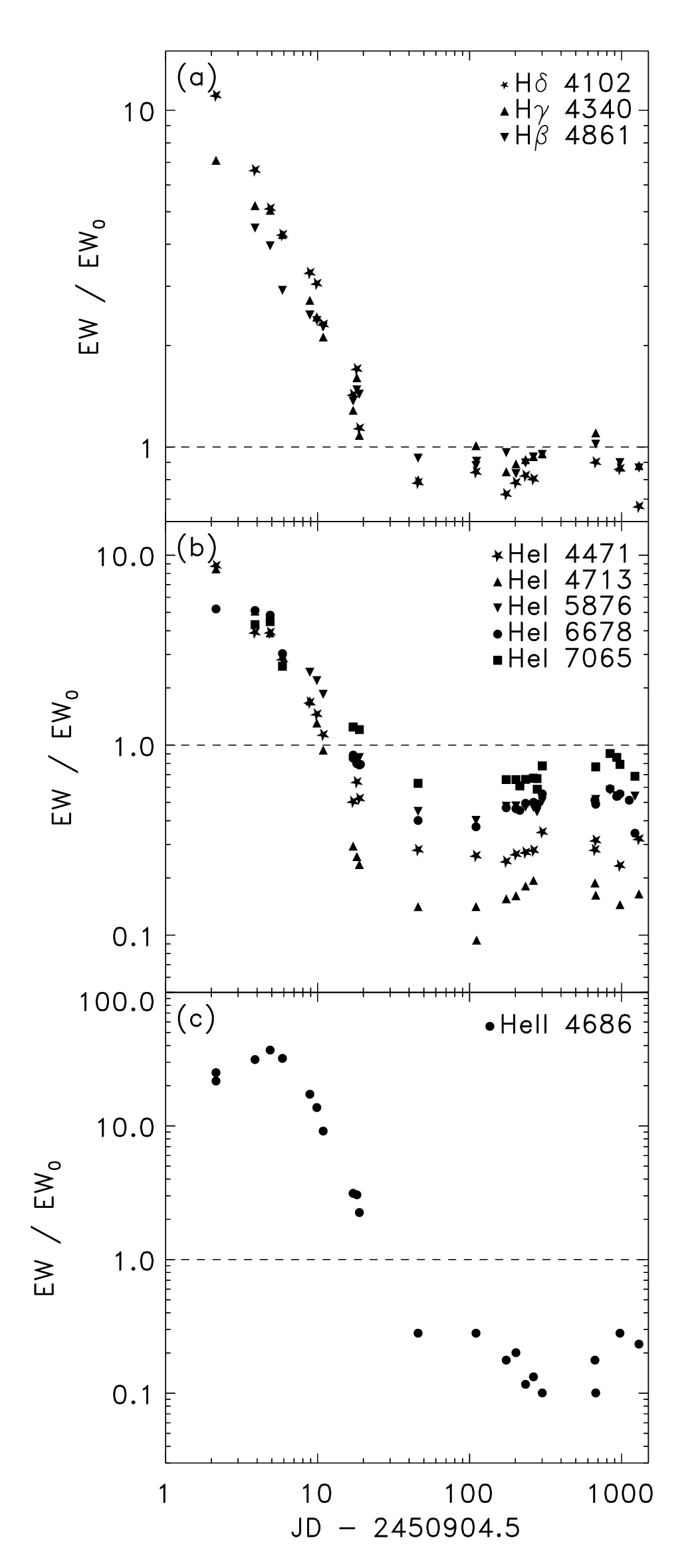} & 
\includegraphics[width=2in]{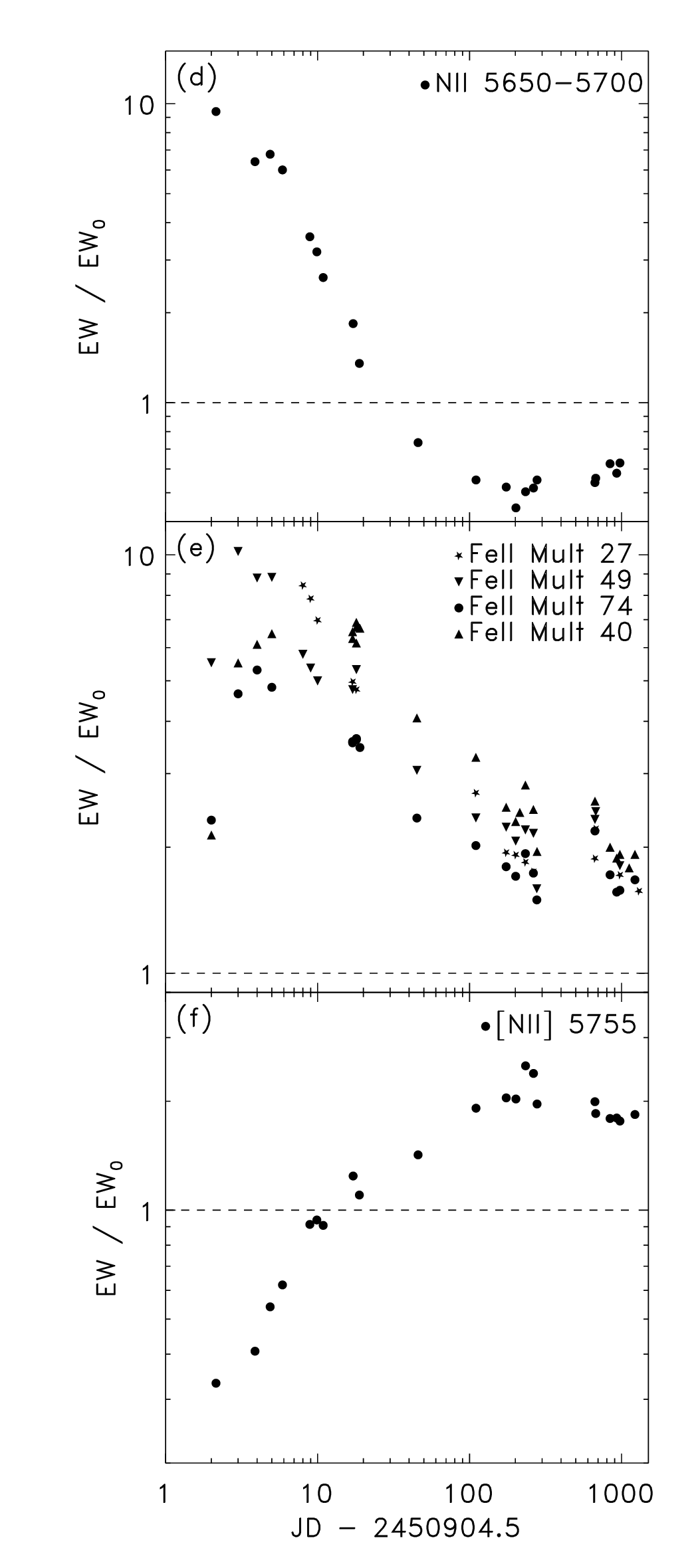} \\
\end{tabular}}
\caption{ Equivalent width evolution of various lines during and after
the outburst.  All equivalent widths have been normalised to
pre-outburst levels, shown as a dashed line, except where noted.  All
points correspond to individual nights except for the 1998 Sept--Dec
FLWO data for which monthly averages have been plotted as the
evolution is slow by this time.  The zero-point of time corresponds to
the peak of the X-ray outburst.  a) Balmer line evolution.  H$\alpha$
is not plotted because it is so strong as to be unreliable in many
spectra.  b) He\,\textsc{i} evolution.  Lines at 4921\,\AA\ and
5015\,\AA\ are omitted as they are blended with strong Fe\,\textsc{ii}
lines in most spectra.  c) The He\,\textsc{ii} 4686\,\AA\ evolution.
d) Evolution of 5650--5700\,\AA\ blend.  This is likely dominated by
N\,\textsc{ii} 5676,5686 at the peak.  e) Evolution of Fe\,\textsc{ii}
lines.  The spectral lines used were: $\lambda\lambda$4233, 4303,
4351, 4385, 4416 (multiplet 27); $\lambda\lambda$6432, 6516 (multiplet
40); $\lambda\lambda$5197, 5234, 5275, 5316, 5325, 5425 (multiplet
49); and $\lambda\lambda$6238, 6248, 6416, 6456 (multiplet 74).  For
each multiplet the sum of equivalent widths has been taken.  For
multiplets 40, 49 and 74 this was normalised to the equivalent sum
before the outburst.  These multiplets are then in approximate
agreement by the end of the period observed (around day 1000), so
multiplet 27, for which no reliable pre-outburst measurements are
possible, was also arbitrarily normalised to agree with them.  The
rise between the first few observations is definitely real, being seen
in all 12 lines for which measurements could be made.  f) Evolution of
[N\,\textsc{ii}] 5755\,\AA\ (filled circles).}
\label{MultiFig}
\end{figure*}

The spectrum of \object{CI~Cam} has long been known to be rich in
emission lines.  In outburst, this is even more dramatic, with most
lines showing larger equivalent widths, and Balmer and He\,\textsc{i}
lines completely dominating their regions of the spectrum (see
Fig.~\ref{MasterSpecFig}).  A similarly rich outburst spectrum was
seen in the infrared (Clark et al.\ \cite{Clark:1999a}).  We now
consider the evolution of selected line strengths and profiles before,
during and after the outburst.  We use all the data described earlier
together with some measurements reported by Wagner et al.\
(\cite{Wagner:1998a}) and Orlandini et al.\ (\cite{Orlandini:2000a}).
Equivalent width lightcurves for a representative range of lines are
collated in Fig.~\ref{MultiFig}; these are discussed in turn in the
following subsections.  No statistical errors have been indicated as
the dominant uncertainty is systematic due to contamination from other
lines and the uncertain continuum level in such a rich spectrum.
Consequently the best indicator of the reliability of a lightcurve is
the scatter of the plotted points.

\subsection{Hydrogen and helium lines}
\label{HydrogenSection}
During outburst H\,\textsc{i}, He\,\textsc{i}, and He\,\textsc{ii}
lines all brighten dramatically.  The increase is much more dramatic
than in the continuum, so equivalent widths rise.  The rising phase is
not observed, except possibly in He\,\textsc{ii} 4686\,\AA\ where the
earliest observations suggest that the equivalent width peaked around
5 days after the X-ray outburst.  Comparison with photometry obtained
at this time indicates that this arises because during this period the
He\,\textsc{ii} flux was approximately constant while the continuum
decayed.  For the H\,\textsc{i} and He\,\textsc{i} lines, the fact
that these lines are already decaying within two days of the X-ray
outburst indicates that they originate in a region that can respond to
the outburst in less than two days.

There are clearly major differences between the H\,\textsc{i} and
He\,\textsc{i} behaviours.  All the Balmer lines appear to return to
approximately their pre-outburst levels within $\sim30$\,days, but the
He\,\textsc{i} lines appear to drop to a factor of 2--10 below this.
This effect has been noted previously by Barsukova et al.\
(\cite{Barsukova:1998a}), Orlandini et al.\ (\cite{Orlandini:2000a})
and Jaschek \& Andrillat (\cite{Jaschek:2000a}) and is clearly real as
can be seen in Fig.\ \ref{MasterSpecFig} where spectra before and
after the outburst are compared.  Further, there appear to be
systematic differences between He\,\textsc{i} lines which are only
manifested after the outburst: during outburst the He\,\textsc{i} EW
ratios are similar to those before the outburst, but afterwards they
can differ, with He\,\textsc{i} 7065\,\AA\ closest to the pre-outburst
level and He\,\textsc{i} 4713\,\AA\ furthest below it.  These
differences are large, corresponding to a change by a factor of three
in the 4713\,:\,6678 ratio for example, so are not simply due to
changes in the continuum shape.  This change in line ratios suggests
changes in the physical conditions in the emitting gas; temperatures
and/or densities; rather than abundance changes (e.g.\ due to ejection
of material as suggested by Orlandini et al.\ \cite{Orlandini:2000a}).
An increase in the 6678\,:\,4471 ratio would be expected from a
decrease in temperature, for example (e.g.\ Osterbrock
\cite{Osterbrock:1974a}).  Changes in the optical depth could also
change ratios, and an increase in the 7065\,:\,4471 ratio could result
from an increase in the optical depth (e.g.\ Osterbrock
\cite{Osterbrock:1974a}).  It is unfortunately impossible to be more
quantitative for such optically thick NLTE lines.  Whatever changes
are involved, they should leave the Balmer lines essentially
unaffected.

\begin{figure}
\resizebox{\hsize}{!}{\includegraphics[angle=90]{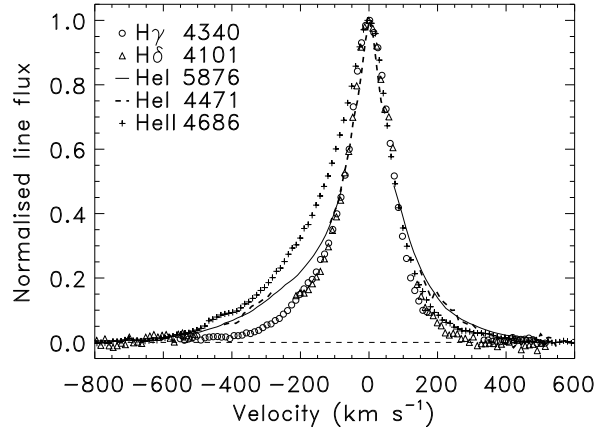}}
\caption{Profiles of the stronger hydrogen and helium lines in
outburst based on WHT/UES data from 1998 April 9.  The profiles have
been rebinned by a factor of four for clarity.  Regions contaminated
by other lines have been omitted.}  
\label{OutburstProfileFig}
\end{figure}

We show a selection of H\,\textsc{i}, He\,\textsc{i}, and
He\,\textsc{ii} line profiles from outburst WHT/UES data in
Fig.~\ref{OutburstProfileFig}.  Robinson et al.\
(\cite{Robinson:2002a}) have already discussed the profiles seen in
outburst.  The WHT/UES data were obtained a few days earlier than
those of Robinson et al.\ (\cite{Robinson:2002a}) but are quite
similar.  They argue that the line profiles, or at least the wings,
are kinematic in origin and hence that emission seen to
$\sim2500$\,km\,s$^{-1}$ indicates the velocity of the hydrogen and
helium emitting material outflowing from the sgB[e] star.  There are
alternative possibilities to be considered, however.  The broad wings
to the lines are most prominent during outburst so may be associated
with material ejected from the accreted compact object rather than
from the sgB[e] star.  Given that these lines are likely formed in
regions of high optical depth (see Sect.~\ref{OxygenSection}),
incoherent Thomson scattering may also broaden the lines
significantly; this has been proposed for other sgB[e] stars (Zickgraf
et al.\ \cite{Zickgraf:1986a}).  Scattering would be expected to be
symmetric, but the observed profiles can be interpreted as the sum of
a broad blue shifted component and a narrower component at rest
(Robinson et al.\ \cite{Robinson:2002a}).  These authors obtained a
satisfactory fit to the profiles with a double Gaussian, with the a
narrow rest component (FWHM 50--85\,km\,s$^{-1}$) and a broad
component (FWHM $\sim160$\,km\,s$^{-1}$) blue-shifted by $\sim
60$\,km\,s$^{-1}$.  We find similar results from our WHT/UES spectra,
although we do not obtain a satisfactory fit with a double Gaussian
model to the He\,\textsc{i} lines.  This is likely due to the
inadequacy of a Gaussian fit rather than an intrinsic asymmetry.  A
double Voigt profile fit, for example, gives extra freedom while still
using symmetric components and works much better.  There is some
sensitivity of the fit parameters to whether a Gaussian or Voigt
profile is used, even when both fit well.  For all fits to all lines
we obtain blue shifts of 40--110\,km\,s$^{-1}$, and FWHM of
60--90\,km\,s$^{-1}$ and 140--230\,km\,s$^{-1}$ for the narrow and
broad components respectively, consistent with the estimates of
Robinson et al.\ (\cite{Robinson:2002a}).  Only the blue shift,
40--110\,km\,s$^{-1}$, needs to be kinematic and we then have much
lower velocities in the hydrogen and helium emission regions, more
comparable to the velocities inferred for metallic lines.  It remains
possible that the width of the broad components is kinematic however.

\begin{figure}
\resizebox{\hsize}{!}{\includegraphics[angle=90]{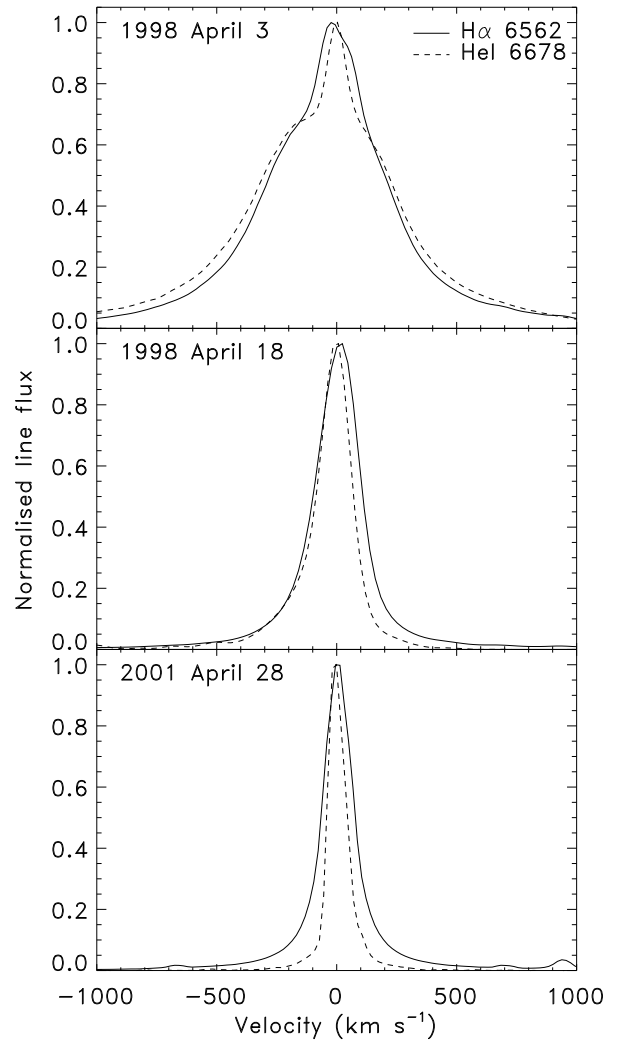}}
\caption{Changes in the profiles of H$\alpha$ and He\,\textsc{i}
6678\,\AA\ from outburst to quiescence.  There is clearly a dramatic
change in the width of the lines, and the early profiles also show a
two component structure.}
\label{FLWOProfileFig}
\end{figure}

During the outburst, large changes are seen in these line profiles.
This is illustrated for H$\alpha$ and He\,\textsc{i} 6678\,\AA\ in
Fig.~\ref{FLWOProfileFig}.  At the earliest epoch (1998 April 3) both
lines are extremely broad and show structure not discernible at later
times.  Emission is clearly seen up to 2500\,km\,s$^{-1}$ in both
lines, and possibly extends to $\sim5000$\,km\,s$^{-1}$ in the blue
wing of H$\alpha$, although this is contaminated by other lines.  The
H$\alpha$ profile in fact looks remarkably like that of
\object{Hen~S134}, a near pole-on sgB[e] star in the LMC (Zickgraf et
al.\ \cite{Zickgraf:1986a}).  Both show a low-velocity blueshifted
`notch', although the overall width of the H$\alpha$ line is much
larger for \object{CI~Cam}.  For \object{Hen~S134}, Zickgraf et al.\
(\cite{Zickgraf:1986a}) suggested that the notch was due to an
unresolved absorption feature.  That might also be true for
\object{CI~Cam}, but the He\,\textsc{i} profile suggests an
alternative explanation.  The latter shows inflections in both the
blue and red wings.  In fact this line clearly suggests two distinct
components: a narrow line at rest and a broad blue-shifted line; the
H$\alpha$ profile may arise the same way, but with a broader rest
component.  This is of course exactly the decomposition that was
applied successfully to later spectra and suggests that rather than
using such a two-component fit as a convenient parameterisation of an
asymmetric profile, we can actually take it more literally: there
really are two distinct emission regions.  Furthermore, fitting the
same two component model to the 1998 April 3 He\,\textsc{i} profile
gives very similar results to the later spectra with a component
separation of $\sim50$\,km\,s$^{-1}$ and a narrow component FWHM of
$\sim75$\,km\,s$^{-1}$. The main difference is that the broad
component in the earlier spectrum is much broader with a FWHM
$\sim750$\,km\,s$^{-1}$.  In fact, there is a hint that the narrow
component of H$\alpha$ on April 3 may itself be more complex.  The
distortion of the peak may indicate that it contains two unresolved
components.  This structure is repeatable over all of the individual
spectra, obtained on this night, and the line peak was not saturated.

\begin{figure}
\resizebox{\hsize}{!}{\includegraphics[angle=90]{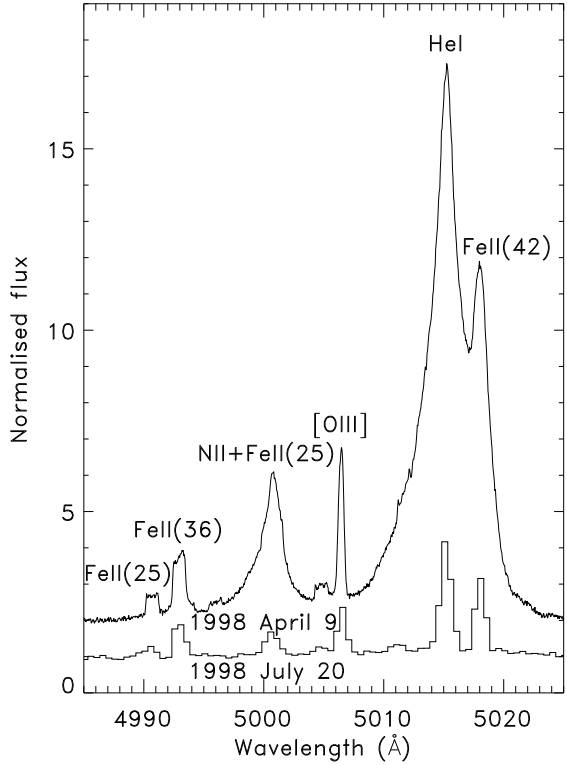}}
\caption{Change in the width of the He\,\textsc{i} 5015\,\AA\ line between
outburst and post-outburst phases.  Note the similar behaviour of the
N\,\textsc{ii} 5001\,\AA\ line.  The upper spectrum has been offset by one
unit vertically.}
\label{HeIWidthFig}
\end{figure}

\begin{figure}
\resizebox{\hsize}{!}{\includegraphics[angle=90]{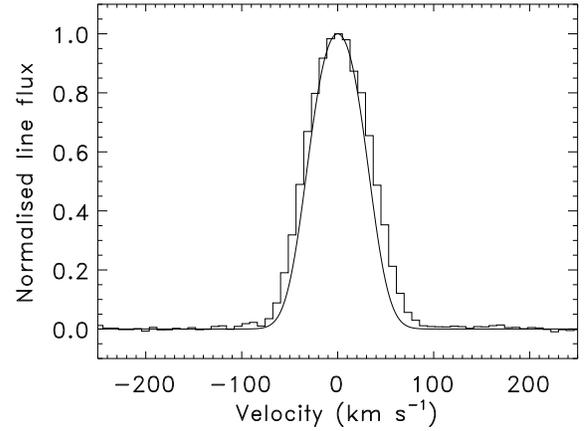}}
\caption{H$\delta$ profile from 2000 December 1.  The underlying
absorption (see Sect.~\ref{CompanionSection}) is much broader than
this, so provides an effectively flat background which has been
subtracted.  The histogram is the data, the smooth line is a square
profile extending to $\pm 32$\,km\,s$^{-1}$ broadened by
14\,km\,s$^{-1}$.  The model profile has also had instrumental
broadening applied, although this is a smaller effect.}
\label{HDeltaProfileFig}
\end{figure}

In quiescence, the weaker hydrogen and helium lines become narrower,
approaching the width of the iron lines (see
Figs.~\ref{FLWOProfileFig}--\ref{HDeltaProfileFig}).  These lines can
never be expected to be as narrow as the iron lines as the thermal
widths will be larger for lighter elements.  Robinson et al.\
(\cite{Robinson:2002a}) observed broadening of 3.1\,km\,s$^{-1}$ for
the Fe\,\textsc{ii} lines, and also estimated a temperature of the
iron emission region of 8000\,K, corresponding to a thermal component
of broadening of 1.9\,km\,s$^{-1}$.  The latter will correspond to
$\sim14$\,km\,s$^{-1}$ for hydrogen lines.
Fig.~\ref{HDeltaProfileFig} shows the observed H$\delta$ profile
(which should be less affected by optical depth effects than stronger
lines) from 2000 December 1, well after the H\,\textsc{i} lines had
stabilised after the outburst.  We have constructed a simple model
profile by taking a square topped profile extending to $\pm
32$\,km\,s$^{-1}$ (the model Robinson et al.\ \cite{Robinson:2002a}
used for the metallic profiles) and broadened it by
$\sim14$\,km\,s$^{-1}$.  This model profile is somewhat narrower than
the observed H$\delta$ profile, but the difference is not dramatic, so
it is plausible that the underlying H\,\textsc{i} kinematics are
similar to the Fe\,\textsc{ii} lines.  Some extra broadening may arise
from a higher temperature in the hydrogen emission region and/or
radiative transfer effects.

It appears that He\,\textsc{ii} 4686\,\AA\ is still detectable after
the outburst, as noted by Barsukova et al.\ (\cite{Barsukova:2002a}).
A close examination of the pre-outburst OHP spectra reveals that it
was also present then at a {\em higher} level, albeit still weak,
indicating that it is not simply a remnant of the outburst.  This is
somewhat surprising and another contaminating line cannot be ruled
out.  Such a line would, however, have to share the property of the
He\,\textsc{i} and N\,\textsc{ii} lines of being weaker after the
outburst than before, which is not consistent with Fe\,\textsc{ii}
line behaviour, for example.  At least one other sgB[e] star shows
He\,\textsc{ii} emission (the luminous B0 star \object{Hen~S134};
Zickgraf et al.\ \cite{Zickgraf:1986a}), however, so this is not
unprecedented, and is further evidence that \object{CI~Cam} is among
the hottest sgB[e] stars.

\subsection{The Bowen blend}
\begin{figure}
\resizebox{\hsize}{!}{\includegraphics[angle=90]{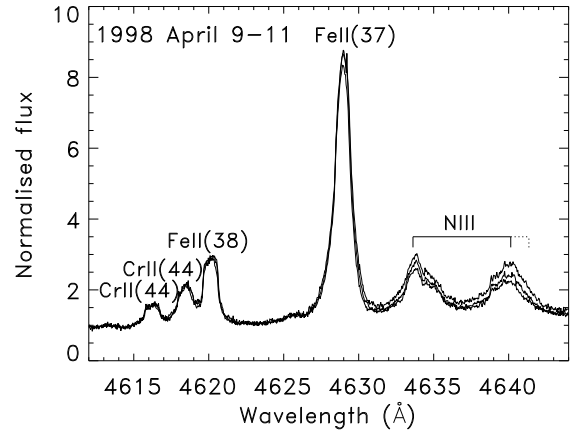}}
\caption{N\,\textsc{iii} Bowen fluorescence lines seen in outburst.
Three normalised spectra are plotted corresponding to the three nights
observed.  The only major changes are in the Bowen features which
become systematically fainter with time.  The three N\,\textsc{iii}
components are indicated, although the 4642\,\AA\ line, indicated by
the dashed line, is not clearly visible and may be weaker and/or
unresolved.}
\label{BowenFig}
\end{figure}

A common feature of the spectra of X-ray binaries is the Bowen blend,
a mixture predominantly of N\,\textsc{iii} and C\,\textsc{iii} lines
spanning $\sim4635$--4650 (McClintock, Canizares \& Tarter
\cite{McClintock:1975a}).  This feature does also appear to be seen in
\object{CI~Cam}, but can only be discerned in the WHT/UES outburst
spectra.  These data are shown in Fig.~\ref{BowenFig}.  Features are
seen corresponding to first two of the N\,\textsc{iii} 4634, 4641,
4642\,\AA\ lines; the third may be present but unresolved from
4641\,\AA.  These features appear broader than Fe\,\textsc{ii} and
Cr\,\textsc{ii} lines in the same region and are clearly declining
much faster.  It therefore seems likely that these are the
N\,\textsc{iii} Bowen fluorescence lines.  The adjacent echelle order
also appears to show the C\,\textsc{iii} 4647, 4650, 4651\,\AA\ lines
which show similar properties.  Neither set of lines appear to be
present after the outburst, although the post-outburst data are of
lower resolution and quality.

\subsection{The 5650--5700\,\AA\ blend}
\begin{figure}
\resizebox{\hsize}{!}{\includegraphics[angle=90]{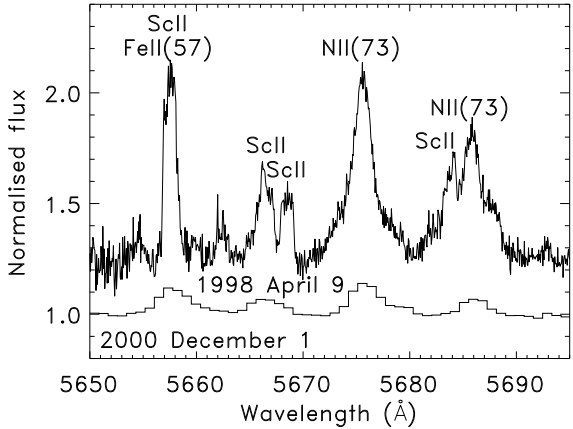}}
\caption{The 5650--5700\,\AA\ blend.  This appears to be a mix of
N\,\textsc{ii}, Sc\,\textsc{ii} and Fe\,\textsc{ii} lines.  In the
earliest outburst spectra the blend is unresolved, but it is clear
that the N\,\textsc{ii} 5676\,\AA\ line dominates, more so than in the
spectra plotted here.  The N\,\textsc{ii} 5679\,\AA\ line should also
be present; it may be detected in the red wing of the 5676\,\AA\ line.
The upper spectrum has been offset by 0.25 units vertically.}
\label{NIIBlendFig}
\end{figure}

There are several lines between 5650 and 5700\,\AA\ which are
dramatically enhanced during the outburst.  In many spectra these are
blended with other lines in the region which are less variable.  The
blend appears to be composed of N\,\textsc{ii}, Sc\,\textsc{ii}, and
Fe\,\textsc{ii} lines (Fig.~\ref{NIIBlendFig}); the variable ones are
likely allowed N\,\textsc{ii} lines from multiplet 73 at 5676,
5686\,\AA.  The strongest predicted line from this multiplet,
5679\,\AA, is not clearly seen, but may be present at a lower level
and poorly resolved from the wing of the 5676\,\AA\ line.  The
evolution of the total equivalent width of the blend is plotted in
Fig.\ \ref{MultiFig}d and shows a rather similar behaviour to the
He\,\textsc{i} lines, with a lower flux after outburst than before.

\subsection{Near-IR oxygen lines}
\label{OxygenSection}
There are a number of important emission lines of O\,\textsc{i}
detected in the near-IR spectrum.  These were not observed often
enough to construct lightcurves, but they still reveal useful
information.  The strongest is O\,\textsc{i} 8446\,\AA.  This has an
equivalent width of 41\,\AA\ before the outburst and 57\,\AA\ after
it.  It is very strong but as discussed by Grandi
(\cite{Grandi:1980a}) there are several possible reasons for this.
Our observations yield rather similar results to those of Chakrabarty
\& Roche (\cite{Chakrabarty:1997a}) for \object{GX~1+4}, so we come to
similar conclusions: that O\,\textsc{i} emission is produced by
H\,\textsc{i} Ly$\beta$ fluorescence; Clark et al.\
(\cite{Clark:1999a}) came to the same conclusion independently from
the 1.13 and 1.32\,$\mu$m infrared lines.  We find that the
O\,\textsc{i} 7774\,\AA\ line is much weaker than 8446\,\AA\, (${\rm
EW}\sim2.6$\,\AA\ before the outburst and 4.2\,\AA\ afterwards),
corresponding to dereddened flux ratios (7774:8446) after the outburst
of order 0.07--0.10 accounting for uncertainty in reddening.  This
rules out production of 8446\,\AA\ by recombination or collisional
excitation, instead favouring a fluorescence process.  Continuum
fluorescence would also predict O\,\textsc{i} 7254\,\AA\ and
7990\,\AA, neither of which are detectable before or after the
outburst, arguing against this interpretation.  The most likely
explanation is then fluorescence by H\,\textsc{i} Ly$\beta$.  This
process requires a gas which is optically thick to H$\alpha$, and the
ratio of 8446\,\AA\ to H$\alpha$ can be used to estimate the H$\alpha$
escape probability (see Grandi \cite{Grandi:1980a}).  We estimate a
dereddened flux ratio (8446:H$\alpha$) after the outburst of
0.07--0.12 and hence an H$\alpha$ escape probability of
$(1.5-2.5)\times 10^{-4}$, implying a very high optical depth in
H$\alpha$.
\subsection{Metallic lines}
\label{IronSection}
\object{CI~Cam} was dubbed `the iron star' by Downes
(\cite{Downes:1984a}) and this is an appropriate designation.  The
spectrum before, during, and after the outburst is rich in
Fe\,\textsc{ii} lines, in many cases blended in all but the highest
resolution spectra.  It is impractical to show the evolution of all
lines because of both the large number of lines and also blending
problems.  In Fig.\ \ref{MultiFig}e we compile results from four
multiplets covering a range of wavelengths and excitations.  Lines
from each multiplet that cannot be reliably deblended or that are
sampled by few spectra are excluded.  It is clear that the decay from
outburst is extended; it may still be continuing even in the most
recent observations.  This is consistent with the observation that the
iron lines appear stronger in post-outburst spectra than in
pre-outburst ones (if they are still enhanced above their `true'
quiescent level).  Minor differences appear between multiplets, e.g.\
compare multiplets 40 and 49.  There is no systematic trend either
with wavelength or with excitation, however, and the origin of the
difference is unclear.  The rise in the early stages of the outburst
is real, as it involves a difference of a factor of two and is
repeated across 12 lines for which measurements could be made over
1998 April 3--6.  It suggests that the iron EW peaked around day 3--10
of the outburst.

\begin{figure}
\resizebox{\hsize}{!}{\rotatebox{90}{\includegraphics{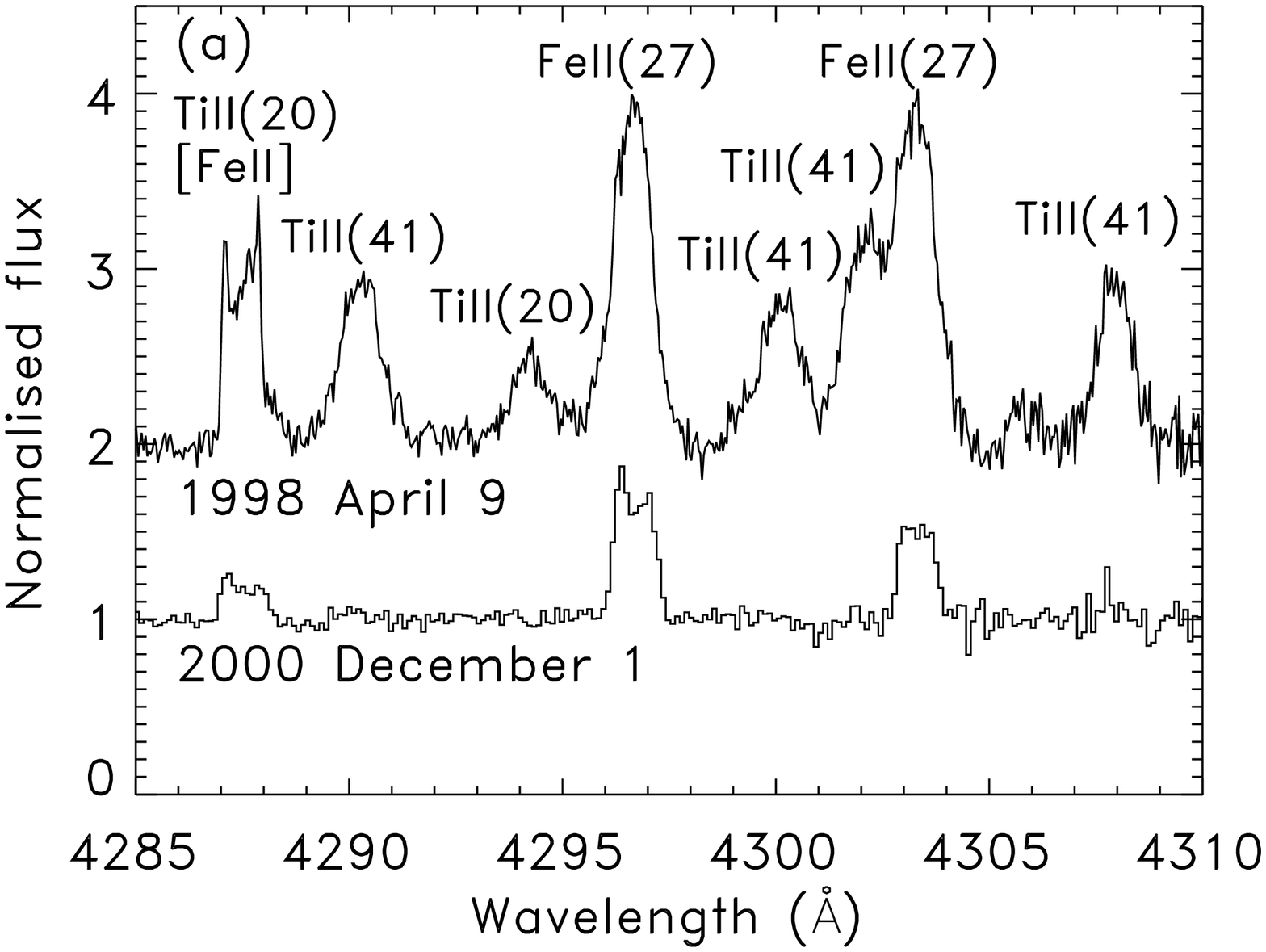}}}
\resizebox{\hsize}{!}{\rotatebox{90}{\includegraphics{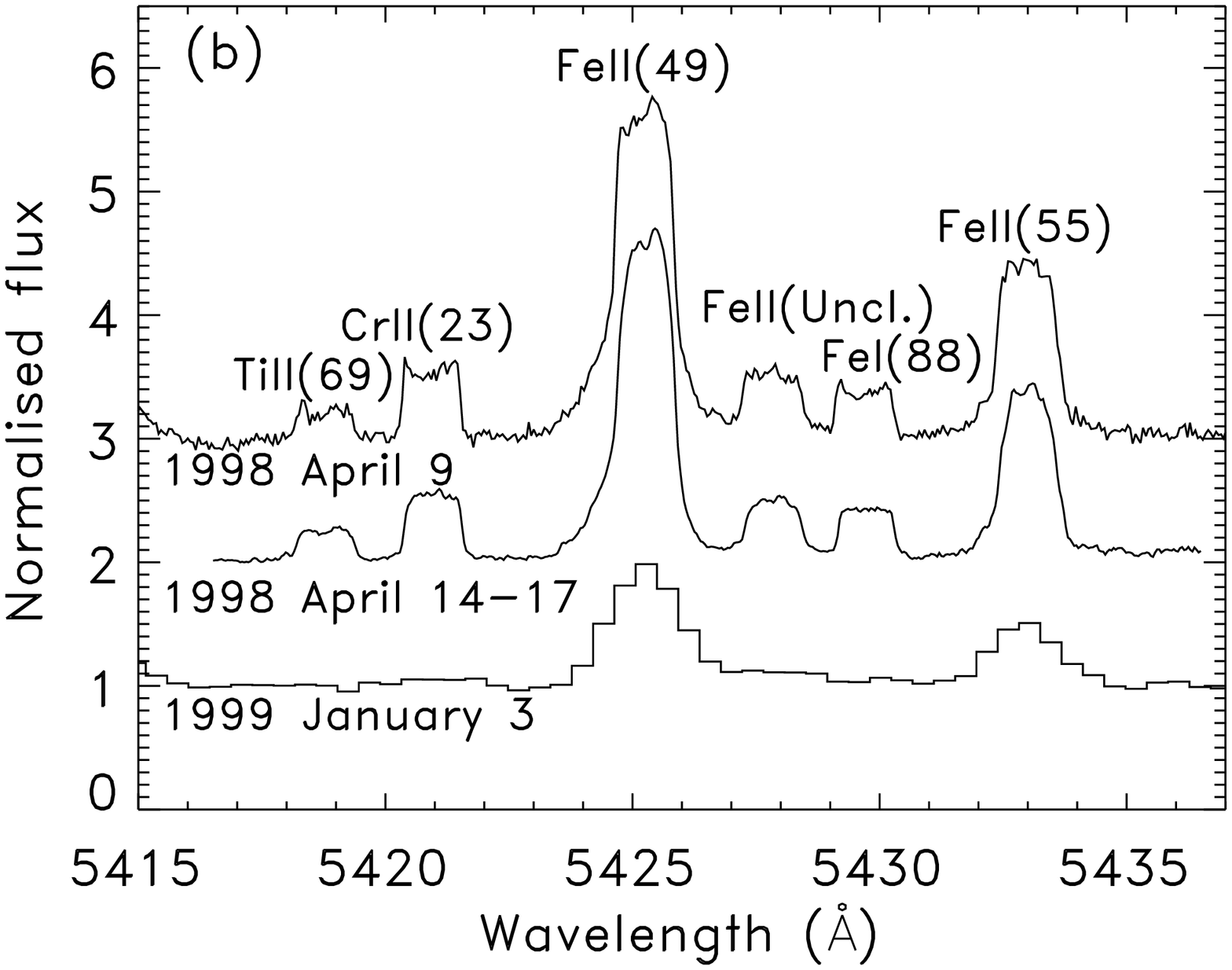}}}
\caption{a) Changes in metallic lines from outburst to quiescence.
Ti\,\textsc{ii} lines are much more variable than Fe\,\textsc{ii}
lines and are virtually undetectable in quiescence.  The
Fe\,\textsc{ii} profiles change, being rounded in outburst but concave
in quiescence (like [Fe\,\textsc{ii}] lines).  Note that the
Fe\,\textsc{ii} line at 4297\,\AA\ in particular does not become
square topped in quiescence.  The upper spectrum has been offset by
one unit vertically.  b) Outburst spectrum of 5415--5437\,\AA\ region.
The spectrum from April 14--17 is reproduced from Fig.~7 of Robinson
et al.\ (\cite{Robinson:2002a}).  Note how the concave tops of the
Cr\,\textsc{ii} and Fe\,\textsc{i} lines in the earliest spectrum
later become flat tops}
\label{TitaniumFig}
\end{figure}

Iron is by no means the only metallic species in the spectrum.  For
example, Robinson et al.\ (\cite{Robinson:2002a}) also identify
permitted lines of Na\,\textsc{i}, Mg\,\textsc{i}, Si\,\textsc{ii},
Ca\,\textsc{ii}, Sc\,\textsc{ii}, Ti\,\textsc{ii}, and Cr\,\textsc{ii}
as well as C, N, and O lines.  The behaviour of these lines can differ
dramatically.  For example Fig.~\ref{TitaniumFig} shows a selection of
Fe\,\textsc{ii} and Ti\,\textsc{ii} lines during and after the
outburst.  The Ti\,\textsc{ii} lines are clearly much more variable
than the Fe\,\textsc{ii} lines and are almost undetectable in
quiescence.  This is puzzling as the Fe\,\textsc{ii} lines shown are
of {\em higher} excitation than the Ti\,\textsc{ii} lines (5.6\,eV for
Fe\,\textsc{ii}; 4.0\,eV for Ti\,\textsc{ii}.), and the ionisation
potential of Fe\,\textsc{ii}, 7.9\,eV, is also higher than that of
Ti\,\textsc{ii}, 6.8\,eV.  Possibly the Ti\,\textsc{ii} lines are
formed in a different region to Fe\,\textsc{ii}.  Cr\,\textsc{ii}
lines may also virtually disappear in quiescence, although all are
quite weak even in outburst.

\begin{figure}
\resizebox{\hsize}{!}{\includegraphics{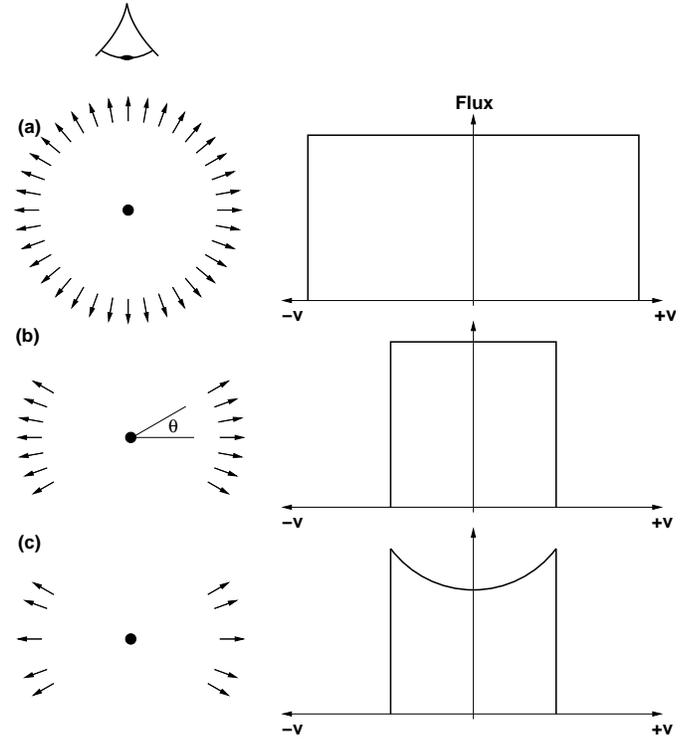}}
\caption{Illustrative optically thin profiles for different outflow
geometries.  The viewing angle is taken to be from the top of the
page, i.e.\ pole-on.  a) A spherically symmetric outflow produces a
square profile extending to $\pm V_{\rm term}$.  b) An equatorial
section of a spherical outflow also produces a square profile (when
viewed pole-on), but narrower, extending to $\pm V_{\rm term}\sin
\theta$. c) If the outflow is non uniform, with less emission closer
to the equatorial plane, then a central depression is created, because
there is less low velocity emission.}
\label{IronModelFig}
\end{figure}

The profiles of the metallic lines have been characterised by Robinson
et al.\ (\cite{Robinson:2002a}) as a square topped profile extending
to $\pm 32$\,km\,s$^{-1}$ and subject to a small thermal broadening.
They attributed this to a uniformly expanding spherical shell
(Fig.~\ref{IronModelFig}a).  We will discuss these lines in the
context of sgB[e] models in Sect.~\ref{BeCompSection}, but here will
note a complication and an alternative explanation.  The complication
is that while a square topped profile clearly provided a good
description of the profiles at the time of the Robinson et al.\
(\cite{Robinson:2002a}) observations, this is not adequate at other
times.  Fig.~\ref{TitaniumFig} shows two segments of our 1998 April 9
WHT/UES outburst spectrum, together with post-outburst counterparts.
The latter are at lower resolution, but are still sufficient to show
deviations from the flat topped profile; the Fe\,\textsc{ii}
4297\,\AA\ line clearly shows double peaks, or a central depression.
This is essentially the same as the [Fe\,\textsc{ii}] 4287\,\AA\ line,
but the latter is also double peaked in outburst.  Other lines in this
spectrum show similar profiles, so this is not just due to noise.  The
problem is worse than this, however, as can be seen in
Fig.~\ref{TitaniumFig}b.  This shows our WHT/UES spectrum of the
region used for Fig.~7 of Robinson et al.\ (\cite{Robinson:2002a}).
Two lines in particular, Cr\,\textsc{ii} 5421\,\AA\ and Fe\,\textsc{i}
5430\,\AA, both of which showed flat topped profiles in the data of
Robinson et al.\ (\cite{Robinson:2002a}) are actually double peaked at
our earlier epoch.  Thus the deviation from a flat topped profile
cannot simply be a post-outburst effect, and the profiles appear to
evolve from a double peaked to flat to double peaked form through and
after the outburst.  Consequently the flat topped profiles seem to be
the exception rather than the norm, and the variations seen suggest
some deviation from spherical symmetry.  We suggest that the
underlying symmetry is axial rather than spherical, but that our
viewing angle is along the axis (i.e.\ we see the star pole-on).  If
we view any equatorial section of a spherical outflow pole-on then the
profile will be rectangular, with narrower equatorial outflows
producing narrower profiles (Fig.~\ref{IronModelFig}b).  If emission
from the equatorial outflow varies with latitude then variations from
a flat top would be seen, and a central depression would imply less
emission near the equatorial plane (Fig.~\ref{IronModelFig}c).

\subsection{Forbidden lines}
\label{ForbiddenSection}
A number of forbidden transitions are observed in [O\,\textsc{i}],
[O\,\textsc{iii}], [N\,\textsc{ii}] and [Fe\,\textsc{ii}].  These are
a rather heterogeneous selection of lines spanning a range of
ionisation stages and showing very diverse behaviours.  As already
noted by Robinson et al.\ (\cite{Robinson:2002a}), [N\,\textsc{ii}]
and [O\,\textsc{iii}] lines are extremely narrow, more so than any
permitted lines (e.g.\ Fig.~\ref{HeIWidthFig}).  [O\,\textsc{i}] and
[Fe\,\textsc{ii}] lines, however, exhibit square-edged profiles
similar to the Fe\,\textsc{ii} lines, although typically with a more
pronounced central depression (e.g.\ Fig.~\ref{TitaniumFig}a).  In the
context of the equatorial outflow geometry suggested for the
Fe\,\textsc{ii} lines, a deeper central depression would imply less
emission close to the equatorial plane.  This makes sense, as the
density is expected to be higher there and so forbidden lines may well
only be seen from higher altitude, lower density, layers.

The line changes with time differ between species too.  Unfortunately,
many of the lines are weak and/or blended with other lines,
principally of Fe\,\textsc{ii}.  The only line well suited to
quantitative analysis is [N\,\textsc{ii}] 5755\,\AA, as it is strong
and unblended.  The equivalent width evolution of this line is shown
in Fig.\ \ref{MultiFig}f.  The early evolution ($\la10$\,days) is
roughly consistent with a constant line flux, with the equivalent
width increasing as the continuum fades.  Comparison with outburst
photometry (Clark et al.\ \cite{Clark:2000a}), after correcting for
the increased contribution of the lines to broad band flux
(Section~\ref{LineContributionSection}), indicates that this outburst
line flux is comparable to that seen in 1998 January.  Barsukova et
al.\ (\cite{Barsukova:2002a}) note that the line flux does increase
modestly around 50--250\,days after outburst, then decays again.  The
decay can clearly be seen in Fig.\ \ref{MultiFig}f, although the rise
is hard to distinguish from that due to the decreasing continuum flux
without flux calibrated spectra.

The [O\,\textsc{iii}] 5007\,\AA\ line also exhibits a narrow profile
but is clearly stronger in outburst than after it
(Fig.~\ref{HeIWidthFig}).  The 4959\,\AA\ line behaves in a similar
way.  These lines are difficult to measure in most spectra as they are
weak and blended, but it is clear that the outburst flux is much
greater than pre-outburst and that after the outburst they decline,
but like [N\,\textsc{ii}], they do not appear to have dropped to the
pre-outburst level.  For example, for [O\,\textsc{iii}] 5007\,\AA, the
pre-outburst EW is difficult to measure but $\sim0.25$\,\AA.  On 1998
April 11 it had an EW of $2.0$\,\AA\ and by 1998 July 20 this had
dropped to $1.1$\,\AA.  As late as 2001 October 23 it was still $\ga
0.8$\,\AA.  The decline in EW from 2000 April to July corresponds to a
decrease of about a factor of three in line flux.  Since the
[O\,\textsc{iii}] lines have very long recombination times, this
implies that significant collisional de-excitation must be occurring,
possibly when the radio ejecta reach the line formation region.
Assuming an expansion velocity of $\sim5000$\,km\,s$^{-1}$, then a
reduction of a factor of three within $\sim100$\,days implies that
most of the [O\,\textsc{iii}] emission originates within
$\sim300$\,AU.

[Fe\,\textsc{ii}] lines also strengthen moderately during outburst
(e.g.\ Fig.~\ref{TitaniumFig}a) and [O\,\textsc{i}] lines increase
enormously, being almost undetectable before and after the outburst.

%%%%%%%%%%%%%%%%%%%%%%%%%%%%%%%%%%%%%%%%%%%%%%%%%%%%%%%%%%%%%%%%%%%%%%%%%%%%%%
%
\section{Short term spectral variability}
\label{Rapid}
There have been some claims of short term variability in
\object{CI~Cam} during outburst.  At X-ray energies Frontera et al.\
(\cite{Frontera:1998a}) found with {\it Beppo-SAX} that the
0.5--1.0\,keV lightcurve showed significant variations on $\sim100$\,s
timescales on 1998 April 9--10.  They found no significant variation
from a smooth decay in the simultaneous 1.5--10\,keV data however, nor
were any variations seen on April 3.  Ueda et al.\ (\cite{Ueda:1998a})
examined {\it ASCA} data from 1998 April 3--4.  They also found that
there was no variability above 1\,keV, but that soft flares were seen
below 1\,keV on timescales of a few hours.  Belloni et al.\
(\cite{Belloni:1999a}) examined {\it RXTE} data spanning April 1--9
and found no evidence for any variation other than a smooth decay at
any time.  The short term X-ray variability thus appears confined to
flares in the soft ($\la 1$\,keV) band.

Optical variability appears sporadic too.  Frontera et al.\
(\cite{Frontera:1998a}) also found evidence for 0.3\,mag optical
flickering on hour timescales on April 6, but not in later
observations spanning April 10--26.  Clark et al.\
(\cite{Clark:2000a}) examined photometry from April 13 and 19 and
found no variations with amplitude greater than 1\,percent.

\begin{figure}
\resizebox{\hsize}{!}{\includegraphics[angle=90]{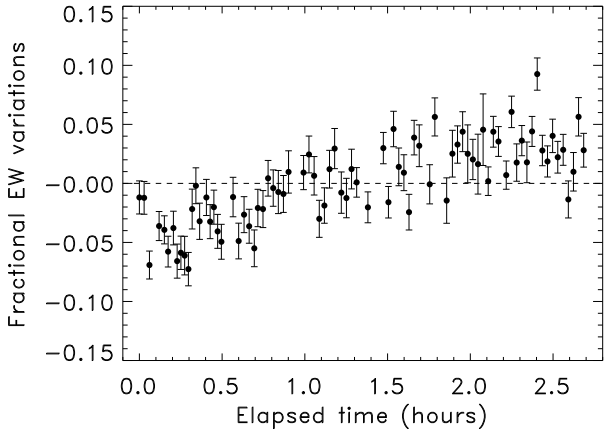}}
\caption{Change in the strength of the He\,\textsc{i} 6678\,\AA\ line
in McDonald spectra from 1998 April 20.  Spectra were normalised to a
flat continuum which was then subtracted.  Integrated line fluxes are
plotted as fractional variations about the mean.  Error bars are
formal errors obtained in extracting one-dimensional spectra and do
not include additional uncertainties that may be introduced by
normalisation of the spectra.}
\label{RapidVarFig}
\end{figure}

The McDonald spectra obtained during the decline from outburst were
taken as a series of short exposures to facilitate studies of line
variability.  Because a relatively narrow slit was used, however,
there are significant slit losses.  Fortunately, as noted above (Clark
et al.\ \cite{Clark:2000a}), there was no detectable short-term
variability in simultaneous $R$ band photometry, so we chose to
normalise each spectrum before extracting emission line lightcurves;
these lightcurves are effectively equivalent widths.  We find no
evidence for variability on timescales from 3\,min to $\sim2$\,hr in
H$\alpha$ or Fe\,\textsc{ii} lines.  Formally we find rms scatters in
the lightcurves of 1.1--1.6\,percent in H$\alpha$ and
0.8--1.5\,percent for a composite of several iron lines.  The one line
that may exhibit short term variability is He\,\textsc{i} 6678\,\AA.
This shows 1.9, 4.9 and 3.5\,percent rms variations on 1998 April 18,
19 and 20 respectively.  On the latter two nights there is clearly a
systematic variation.  To be sure this is not an artifact we have
renormalised the continuum using just the 6648--6663 and
6693--6708\,\AA\ regions and recentered the line in each spectrum by
cross-correlating the line profiles with an average.  None of these
changes affect the result: this line appears to show real variability.
The case is most persuasive on the last night.  There is a clear
overall rise in the linestrength of $\sim10$\,percent over
$<3$\,hours, shown in Fig.\ \ref{RapidVarFig}.  This is in the
opposite direction to the overall decline in the linestrength and much
larger than expected from the changes in the continuum strength alone
(e-folding decay time $\sim24$\,days at this time; Clark et al.\
\cite{Clark:2000a}).  We have also constructed rms spectra for each
night and compared them with the average spectra.  These suggest the
same conclusion as the lightcurves; the only feature that seems to
show significant variability is the He\,\textsc{i} line on the second
and third nights.

We also examined the FLWO data from 1998 April 3 in the same way.  The
rms spectrum shows no variable features other than Telluric bands.
The line strengths for H$\alpha$ and He\,\textsc{i} 6678\,\AA\ show an
rms scatter of 1.8\,percent and 1.2\,percent respectively with no
systematic trend.  We believe this represents a null detection of
variability on this night.
%
%%%%%%%%%%%%%%%%%%%%%%%%%%%%%%%%%%%%%%%%%%%%%%%%%%%%%%%%%%%%%%%%%%%%%%%%%%%%%%
%
\section{Discussion}
\label{Discussion}
\subsection{\object{CI~Cam} as an sgB[e] star}
\label{BeCompSection}

It is clear that the X-ray outburst was associated with dramatic
changes in the spectrum of \object{CI~Cam}.  Having identified it as
an sgB[e] star, however, it will prove useful to first compare it with
the small number of other stars of this class and see how much of its
`unusual' behaviour, is typical of these objects, and what requires
additional input from a compact object.

Robinson et al.\ (\cite{Robinson:2002a}) infer a predominantly
spherical geometry for the outflow from \object{CI~Cam}.  Several
components are suggested.  A cool, low velocity (32\,km\,s$^{-1}$)
wind gives rise to the many iron lines and the square profiles of
these lines suggested that this component is spherical.  Multiple
higher velocity components ($>1000$\,km\,s$^{-1}$) are suggested by
hydrogen and helium lines, and by the UV resonance lines.  This
picture has several problems as noted by Robinson et al.\
(\cite{Robinson:2002a}), principally i) it is unclear how several
near-spherical winds can co-exist with very different velocities and
ionisations; and ii) the quiescent X-ray luminosity is very low for a
compact object continually burrowing through a dense, spherical
outflow.

This picture is also rather at odds with existing models of sgB[e]
stars (Zickgraf et al.\ \cite{Zickgraf:1985a},\cite{Zickgraf:1986a};
Oudmaijer et al.\ \cite{Oudmaijer:1998a}).  The structure inferred
from other systems is of a two-component wind which is far from
spherical.  A rarefied, hot, high velocity wind dominates in the polar
regions and is responsible for the UV lines often seen with P~Cygni
profiles.  This wind is identical to the winds of normal early-type
supergiants.  Other lines are attributed to a cooler, denser wind
concentrated in the equatorial regions that is not present in normal
supergiants.  The nature and origin of this equatorial material remain
unclear, and it is probably quite different to the discs around
classical Be stars.  Can \object{CI~Cam} be fitted into this
framework?

The interpretation of the UV lines seen in \object{CI~Cam} is most
straightforward.  These show broad P~Cygni lines typical of supergiant
winds.  They can be associated with the hot, high velocity, polar
outflow, and we essentially agree with Robinson et al.\
(\cite{Robinson:2002a}) on the interpretation of these lines.

The hydrogen and helium lines are more problematic, however.  In
outburst, these lines are broad and asymmetric, with blue-shifted
emission extending to $\ga2500$\,km\,s$^{-1}$, but with no detectable
absorption components.  Robinson et al.\ (\cite{Robinson:2002a})
associate these lines with a high-velocity, weakly collimated outflow
from the sgB[e] star.  It is unclear where this could be situated,
however: if it originated from the same region responsible for the UV
lines we might expect P~Cygni line profiles, as seen in
\object{HD~87643} (Oudmaijer et al.\ \cite{Oudmaijer:1998a}).  The
latter object may be rather different to other sgB[e] stars with a
higher H\,\textsc{i} column density in the polar wind.  In most sgB[e]
stars, hydrogen and helium emission is instead attributed to the
equatorial component, and hence relatively low velocities should be
involved, as is seen in the iron lines.  In fact, a clue that the
hydrogen, helium and iron lines are all associated with the same
region, or regions, is provided by similarities of the line profiles.
Robinson et al.\ (\cite{Robinson:2002a}) associate Fe\,\textsc{ii}
lines with a very different region to the hydrogen and helium lines,
but they also note that {\em `The stronger [iron] lines have more
rounded profiles, becoming almost Gaussian in shape as the lines
become stronger, and the very strongest lines have an extended blue
wing.'}  Another way to put this is that the strongest lines become
more akin to hydrogen and helium lines.  While the rounded profiles
could indicate radiative transfer effects, the presence of asymmetry
in the Fe\,\textsc{ii} lines suggests that the same rest and
blue-shifted components are present in the iron lines as are seen in
hydrogen and helium profiles.  If the strongest, optically thickest
iron lines look like hydrogen lines, then the corollary is that the
weakest, optically thinnest hydrogen and helium lines should look like
iron lines.  This does seem to be the case, as demonstrated in
Sect.~\ref{HydrogenSection} and Fig.~\ref{HDeltaProfileFig}.

The decomposition of the hydrogen and helium lines into two clear
components suggests that two distinct regions are involved during
outburst.  The narrow rest component dominates in the iron lines and
is also important in the hydrogen and helium lines.  A second broad,
blue-shifted component is present in hydrogen and helium lines, and
weakly detectable in the strongest iron lines.  The narrower, rest
component is likely the same component as seen in quiescence, and in
other sgB[e] stars, originating from the equatorial outflow.  The
profiles seen in H\,\textsc{i} and He\,\textsc{i} are very different
to the Fe\,\textsc{ii} lines, but the former are broadened by larger
thermal widths and incoherent Thomson scattering, as suggested by
Zickgraf et al.\ (\cite{Zickgraf:1986a}) for other sgB[e] stars.  The
strong, broad, blue-shifted component is a feature of the outburst and
may not be associated with the outflow from the sgB[e] star at all.
We will discuss this possibility further in the following section.

Associating the iron lines with the equatorial material conflicts with
the interpretation of their profiles offered by Robinson et al.\
(\cite{Robinson:2002a}).  They argue that the rectangular profiles
allow little deviation from spherical symmetry and no significant
rotational velocity.  As we have discussed in Sect.~\ref{IronSection},
however, large deviations from spherical geometry are possible in one
case: that an equatorial outflow is viewed pole-on.  A pole-on
equatorial outflow can also explain the apparent lack of rotational
velocities, as the {\em projected} rotational velocity could be very
low, and the intrinsic rotational velocity in regions of the disc
where Fe\,{\sc ii} can arise is in any case likely to be very low.

As a further argument for the presence of a spherical low velocity
wind, Robinson et al.\ (\cite{Robinson:2002a}) claim on the basis of
the scaling relation of Bjorkman (\cite{Bjorkman:1998a}) that dust can
condense in this wind and that an equatorial outflow is therefore not
required for the production of dust. The scaling relation of Bjorkman
(\cite{Bjorkman:1998a}) is based on the argument that the density of
wind compressed outflows in sgB[e] stars at radii where the wind is
cool enough for dust to condense is $\sim$equal to those of post-AGB
stars (which do form dust) and therefore that it is possible for dust
to condense in such outflows. However, the scaling relation does not
allow for the fact that the winds in post-AGB stars are significantly
enhanced in species such as Si and C (via dredge up) which form dust
up to a factor of $\sim$10 times more easily relative to the winds
from sgB[e] stars.  Additionally, the mass loss for the sgB[e] star is
calculated from the base density of the outflow at the surface of the
star and the {\em terminal velocity} of the outflow, rather than the
velocity of the outflow {\em at the surface of the star}. This will
also serve to overestimate the density of the outflow at the dust
condensation radius. Combining the 2 arguments we estimate that the
density is likely to be 2--3 orders of magnitude too small to allow
dust to condense (J. Porter, priv.\ comm.). While this does not {\em a
priori} argue against the presence of a quasi-spherical wind resulting
in the Fe\,{\sc ii} line profiles it {\em does} demonstrate that dust
cannot condense in it and that a region of higher density -- whether
an equatorial outflow or dense ejecta -- is required for condensation.
Robinson et al.\ (\cite{Robinson:2002a}) additionally argue that the
presence of a resolved, circular dust shell (Traub, Millan-Gabet \&
Garcia \cite{Traub:1998a}) is consistent with their interpretation,
but we note that if a disc is viewed pole-on then dust in the outer
regions of the disc will produce the same apparent structure.

Forbidden lines are a rather heterogeneous mixture, but broadly divide
into two types.  Narrow lines ([N\,\textsc{ii}], [O\,\textsc{iii}])
are formed in a region of very low bulk, turbulent and/or thermal
velocities.  This probably places them outside the two component
outflow from the sgB[e] star, and they may be a remnant of an earlier
phase of evolution of one of the stars (c.f.\ Oudmaijer et al.\
\cite{Oudmaijer:1998a}).  The [Fe\,\textsc{ii}] lines show profiles
similar to the Fe\,\textsc{ii} lines, but with a more pronounced
central depression.  As discussed in Sect.~\ref{ForbiddenSection},
this suggests that these lines may be formed in the upper layers of
the same equatorial material responsible for Fe\,\textsc{ii};
[O\,\textsc{i}] lines are likely formed in the same region.

To summarise, by analogy to other known sgB[e] stars we suggest that
when the system is not in outburst, the strongest optical permitted
lines (H\,\textsc{i}, He\,\textsc{i}, Fe\,\textsc{ii}, and others) and
some forbidden lines ([Fe\,\textsc{ii}], [O\,\textsc{i}]) arise in an
equatorially concentrated region, whether a quasi-Keplerian disc
(e.g.\ Okazaki \cite{Okazaki:2000a}), a hybrid disc with a slow radial
expansion velocity, or even in a radiation driven disc wind such as is
suggested in the Galactic sgB[e] star \object{HD~87643} (Oudmaijer et
al. \cite{Oudmaijer:1998a}).  UV P~Cygni lines originate from a
hotter, higher velocity polar outflow of much lower density.  The
whole structure is viewed close to pole-on, hence disc lines have
rather low velocities as both rotational and expansion velocities
would be mainly directed in the equatorial plane.  Finally the narrow
forbidden lines ([N\,\textsc{ii}], [O\,\textsc{iii}]) come from a much
more extended and near stationary region.

Our suggested model does depend on viewing the source near pole-on.
There are several independent lines of evidence to support this,
suggesting that \object{CI~Cam} is seen at a lower inclination than
many sgB[e] stars.

\begin{enumerate}
\item
Most sgB[e] stars show low velocity blueshifted absorption in their
Balmer line profiles (Zickgraf et al.\
\cite{Zickgraf:1985a},\cite{Zickgraf:1986a}; Oudmaijer et al.\
\cite{Oudmaijer:1998a}).  This can be interpreted as due to absorption
by the slow equatorial wind component.  No such absorption is seen in
\object{CI~Cam} suggesting that the inclination is low enough that our
line of sight does not pass through the slow wind.
\item
sgB[e] stars are expected to show significant polarisation changes
through H$\alpha$.  The observations of \object{HD~87643} are
qualitatively similar to what is expected for a rotating, expanding
disc (Oudmaijer et al.\ \cite{Oudmaijer:1998a}).  In \object{CI~Cam}
no polarisation changes are seen through the lines (Ikeda et al.\
\cite{Ikeda:2000a}).  In the models of Wood et al.\
(\cite{Wood:1993a}) the polarisation decreases with inclination, so a
lack of line polarisation also supports a low inclination.  In
addition, the outburst observations of Ikeda et al.\
(\cite{Ikeda:2000a}) showed a continuum polarisation consistent with
surrounding field stars, and hence with purely interstellar
polarisation.  This is also as expected for a nearly pole-on disc.
\item
We have argued above that dust cannot form in a spherical shell and
that it must be associated with the outer disc.  The near perfectly
circular resolved infrared image then suggests that this disc must be
viewed pole-on.
\end{enumerate}

It therefore seems plausible that \object{CI~Cam} is a pole-on sgB[e]
star.  Such a model is consistent with the properties of both the
optical and UV spectra.

\subsection{The nature of the outburst}

We now move on to discuss the role of the compact object.  Its nature
obviously has some bearing on this discussion.  The X-ray outburst was
very short, much shorter than typical of soft X-ray transients (SXTs).
This, together with the radio emission from an expanding remnant
(Mioduszewski et al.\ 2002, in preparation) suggests some kind of
explosive event.  Orlandini et al.\ (\cite{Orlandini:2000a}) suggest a
thermonuclear runaway on the surface of a white dwarf is responsible,
but this seems hard to reconcile with the high inferred X-ray
luminosity (Robinson et al.\ \cite{Robinson:2002a}) and the
$\gamma$-ray detection by {\it CGRO}/BATSE (Belloni et al.\
\cite{Belloni:1999a}).  It seems more likely that the outburst
involved a brief burst of supercritical accretion onto a neutron star
or black hole, resulting in ejection of much of the accreted material.
This accounts for the observed radio ejecta and possibly the broad
components of the hydrogen and helium lines.  A supercritical
accretion model could also involve a large enough optical depth of
scattering material around the X-ray source to smear out any short
timescale variability.

One possible cause for such a large burst of accretion would be the
passage of the compact object through the equatorial plane.  We can
attempt to estimate the accretion rate, although this will be very
uncertain as the orbital parameters are unknown and the physical
conditions in sgB[e] star outflows are not well understood.  We assume
stellar parameters typical of hot, luminous sgB[e] stars (Zickgraf et
al.\ \cite{Zickgraf:1986a}) and adopt a representative 30\,year orbit
of a 10\,M$_{\odot}$ black hole around a 60\,M$_{\odot}$ sgB[e] star
with an eccentricity of $e\sim0.9$.  The periastron distance is then
about 10--20\,R$_{\rm star}$ and the periastron velocity
170\,km\,s$^{-1}$.  In the model of Oudmaijer et al.\
(\cite{Oudmaijer:1998a}) for \object{HD~87643}, the equatorial density
at 10\,R$_{\rm star}$ is $\sim2.5\times 10^{-12}$\,g\,cm$^{-3}$.  This
model may not be correct, and \object{CI~Cam} is probably more massive
and luminous than \object{HD~87643}, but this at least indicates the
kind of equatorial density which is considered plausible.  The
velocity of the compact object relative to the equatorial material
will be dominated by its orbital motion, so the Bondi accretion rate
is then predicted to be $\sim6\times10^{21}$\,g\,s$^{-1}$ or
$\sim400$\,$\dot{\rm M}_{\rm Edd}$ (Bondi \cite{Bondi:1952a}).  Given
the uncertainties mentioned above, this is a very approximate
estimate, and could be at least an order of magnitude off.  Even
allowing for this large uncertainty, the accretion rate in this
scenario can therefore be expected to be extremely high, due to the
high density and low velocity of the equatorial material, and hence a
supercritical accretion episode is possible.  In contrast, when the
compact object is out of the plane the density is lower by a factor of
$10^4$ (in the model of Oudmaijer et al.\ \cite{Oudmaijer:1998a}) and
the relative velocity is higher, since the high latitude outflow moves
much faster.  The Bondi accretion rate is then expected to be much
lower, $\sim8\times10^{-5}$\,$\dot{\rm M}_{\rm Edd}$.  If the
accretion efficiency remained high, $\eta\sim0.1$, then
\object{CI~Cam} should still be a relatively bright X-ray source, with
$L_{\rm X}\sim10^{35}$\,erg\,s$^{-1}$.  However, at these low
accretion rates then the flow could be expected to become advective as
proposed for other quiescent black hole candidates (Narayan, Garcia \&
McClintock \cite{Narayan:2001a} and references therein), and the
accretion efficiency would then be lower,
$\eta\sim10^{-2}$--$10^{-4}$, implying $L_{\rm X} \sim
10^{32}$--$10^{34}$\,erg\,s$^{-1}$ as observed (Robinson et al.\
\cite{Robinson:2002a}).  This model therefore avoids the problem that
Robinson et al.\ (\cite{Robinson:2002a}) had, that a compact object
burrowing through a dense spherical outflow should be persistently
bright.

The passage of a compact object through the equatorial material, and
the ejection of a lot of material from near it could affect the
equatorial region significantly, so changes in the line spectrum from
this region would be expected.  We cannot offer an exact mechanism for
this, however, and several factors may be involved: the tidal effect
of the compact object passage; X-ray heating; and the interaction of
the expanding radio remnant with the equatorial flow.  The rapid
response of the lines to the outburst, peaking within a few days of
the X-rays, constrains the tidal effect, as a tidally triggered disc
outburst would be expected to proceed on a viscous timescale, if the
disc is Keplerian (i.e.\ thin and rotationally supported, with small
inflow or outflow velocity).  This is likely to be very long, hundreds
to thousands of days.  It is far from clear that sgB[e] discs are
Keplerian, but other timescales, e.g.\ the outflow time from the
stellar surface to $\sim50$ stellar radii, are likely to be similar.
The extended decay of the iron lines may indeed be due to a recovery
of the disc to its pre-outburst state on this timescale, but the rise
of the outburst cannot be, so a tidally triggered disc outburst is
unlikely.

The X-ray heating effect will obviously be much faster, and should
rise and fade on a timescale comparable to the X-ray outburst itself,
although could be prolonged by significant cooling or recombination
times.  This is not seen, although a signature of X-ray heating of the
disc may be present in the form of strong 6--7\,keV line emission seen
by {\it SAX} (Frontera et al.\ \cite{Frontera:1998a}), {\it ASCA}
(Ueda et al.\ \cite{Ueda:1998a}) and {\it RXTE} (Revnivtsev et al.\
\cite{Revnivtsev:1999a}; Belloni et al.\ \cite{Belloni:1999a}).  If
this line is attributed to fluorescent iron emission then a large,
cold, optically thick surface is needed covering about half the sky as
seen from the X-ray source (Ueda et al.\ \cite{Ueda:1998a}).  This is
what would be expected if the X-ray source were above the plane of the
equatorial material.  The dramatic enhancement of high excitation
lines, such as He\,\textsc{ii} and N\,\textsc{iii} during outburst
also suggests that X-ray or extreme-UV irradiation may place a role.

The expansion of the radio ejecta can more naturally produce the
timescales observed than either a tidal interaction or X-ray
irradiation can.  Assuming an expansion at $\sim5000$\,km\,s$^{-1}$
(from Mioduszewski et al.\ in preparation, assuming $d\sim5$\,kpc),
this would reach a disc radius of $\sim50$\,stellar radii
($\sim2000$\,R$_{\odot}$) in about 3\,days, so this is consistent with
the relatively short outburst and the rise time of the Fe\,\textsc{ii}
EWs.  The expanding ejecta may also explain the broad blue-shifted
component seen prominently in hydrogen and helium profiles during
outburst, as the inferred expansion velocities are comparable to the
maximum detectable velocities in the emission lines.  This is an
appealing interpretation, as the origin of these ejecta is most likely
the compact object, and hence an overall blue shift can be explained
as the orbital velocity of the compact object at the time of ejection.
The overall blue-shift seen, 50--100\,km\,s$^{-1}$, is reasonable
compared to our estimate of the periastron velocity
$\sim170$\,km\,s$^{-1}$.  It is also sensible that we see a
blue-shift, rather than red, as X-ray spectra taken in the outburst
decay show a strong fluorescent iron line, consistent with reflection
from the equatorial material, and do not show strong absorption; both
indicate that the compact object was likely between us and the
equatorial plane at the time, and hence should have been moving toward
us.

\subsection{The origin of the optical continuum}
\label{ContinuumOriginSection}

The dramatic increase in the optical luminosity of \object{CI~Cam}
during the outburst is also intriguing.  The optical rise was not
observed, but if it was later than the X-rays then the rise time was
fast, $\la2$\,days.  While large optical brightening is typical of
SXTs, it is actually rather surprising in a system containing a hot,
luminous sgB[e] star.  The brightening is such that whatever provides
the additional source of light during outburst must significantly
outshine the sgB[e] star itself.  Relative to the mean pre-outburst
level (Bergner et al.\ \cite{Bergner:1995a}) the brightest magnitudes
reported are brighter by 1.9\,mag ($U$, Apr 3.9; Hynes et al.\
\cite{Hynes:1998a}), 2.1\,mag ($B$, Apr 3.1; Garcia et al.\
\cite{Garcia:1998a}), 2.3\,mag ($V$, Apr 3.1; Garcia et al.\
\cite{Garcia:1998a}), 3.5\,mag ($R$, Apr 2.1; Robinson et al.\
\cite{Robinson:1998a}), 2.4\,mag($I$, Apr 3.8; Clark et al.\
\cite{Clark:2000a}).  These observations are not simultaneous and are
not intended to indicate the spectrum; merely to indicate that the
outburst was dramatic throughout the optical region.  Since the peak
of the outburst was missed in the optical these are actually lower
limits on the outburst amplitude.  While some of the brightening is
due to stronger line emission, not all can be; removing the line
emission from our early spectra only increases $B$ and $V$ by
$\sim0.6$\,mag, so much of the brightening has to be enhanced
continuum.  Fig.~\ref{FluxSpecFig} indicates that on 1998 April 3, the
excess continuum was 2--5$\times$ brighter than the sgB[e] star, and
redder than it.  Where can this extra continuum come from?  Some
possibilities which we discuss below are:

\begin{enumerate}
\item
Heating of the sgB[e] star during the outburst.
\item
Heating of the dusty envelope around the system.
\item
Emission from an accretion disc around the compact object.
\item
Enhanced emission from the circumstellar material around the sgB[e] star.
\item
Emission related to the observed radio ejection.
\end{enumerate}

\begin{figure}
\resizebox{\hsize}{!}{\includegraphics[angle=90]{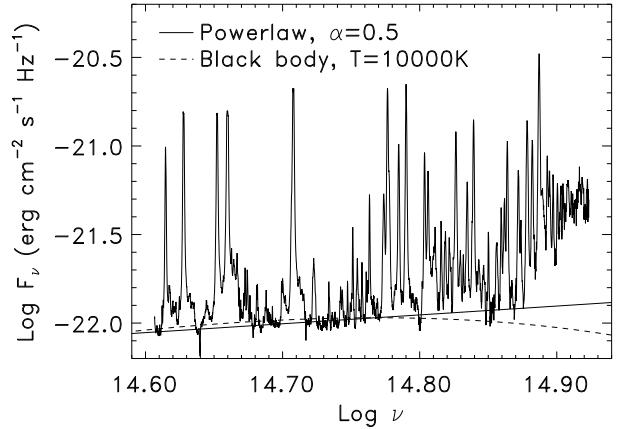}}
\caption{Spectrum of the extra light seen in outburst, obtained by
taking the difference between spectra obtained on 1998 April 3 and
1998 October 29.  This has been dereddened using the Fitzpatrick
(\cite{Fitzpatrick:1999a}) extinction curve assuming $A_V=4.0$.  Two
possible models for the continuum spectrum are also plotted; the power
law corresponds to $F_{\nu} \propto \nu^{1/2}$.}
\label{FluxSpecModelFig}
\end{figure}

To help constrain the possibilities it is useful to examine the
spectrum of the excess light.  Fig.~\ref{FluxSpecModelFig} shows the
difference between the early outburst and post-outburst spectra from
Fig.~\ref{FluxSpecFig}.  Close examination of the outburst spectrum,
and comparison with high resolution spectra taken later in outburst,
indicates that although much of the spectrum is dominated by emission
lines, the lowest points do appear to indicate the continuum.  The
difference spectrum is obviously rather uncertain, but many
possibilities can be ruled out.  If it is a power-law, then it is
relatively flat in $f_{\nu}$.  The power-law model plotted is
$f_{\nu}\propto \nu^{1/2}$.  A completely flat spectrum would also be
possible, but it could not be much redder or bluer than these
possibilities.  A hot black body is ruled out, as indicated by the
fact that the difference spectrum is redder than the sgB[e] spectrum.
A 10\,000\,K black body is shown; a temperature much cooler than this
would also not be consistent with the data.  Similar conclusions can
be drawn by examining the change in $B-V$ during the outburst, from
$\sim0.8$\,mag to $\la1.2$\,mag.  It is obviously difficult to be
precise given the uncertain reddening and large line contribution, but
a very hot or very cold black body, or a very red or very blue
power-law can be ruled out.

This colour information can immediately be used to constrain the
origin of the excess light.  Interpretation 1 can be ruled out as
heating of the sgB[e] star could not make the spectrum redder even if
a $5\times$ increase in brightness were possible.  In this case, the
excess should be extremely blue.  Equally, interpretation 2 can be
ruled out as the excess light is not cool enough.  Heated dust would
dissociate for $T\ga 2000$\,K, but the observed spectrum and change in
$B-V$ are clearly not consistent with such a low temperature.

The brightness of the outburst argues against interpretation 3.  After
correcting for line emission ($\Delta V\sim0.6$\,mag) and extinction
($A_V \sim 4$\,mag), the peak dereddened continuum magnitude is $V\la
5.8$ and the average quiescent value is $\left< V \right> \sim
7.7$\,mag; hence the additional continuum source has $V \la 6.0$.  At
a distance of $\sim5$\,kpc this corresponds to an absolute magnitude
of $M_V \la -7.5$.  This is much brighter than any low mass X-ray
binary (LMXB) ($-5 < M_V < +5$; van Paradijs
\cite{vanParadijs:1994a}), i.e.\ much brighter than other observed
accretion discs around black holes or neutron stars.  If the
additional light were an accretion disc it would then have to be more
luminous than those in LMXBs. Since the spectrum indicates that it is
not extremely hot, it would have to be large.  As the size of LMXB
accretion discs is limited by tidal truncation, a much larger disc in
\object{CI~Cam} is certainly possible.  If the density in the disc
were comparable to the disc densities in LMXBs, however, such a large
disc would provide enough mass to sustain a much longer X-ray outburst
than seen, and a decay timescale of months, more like that seen in
SXTs, would be expected.  Indeed Robinson et al.\
(\cite{Robinson:2002a}) have argued that the very rapid decay
timescale in \object{CI~Cam} indicates that the accretion disc, if
present, must be very small.  This is a very different environment
from that in an LMXB, however, and the disc could be stabilised
against thermal instability by irradiation from the supergiant at much
lower densities than required in an LMXB, resulting in a lower disc
mass.  The brightness remains a problem, however, as the disc
luminosity would be comparable to that of the supergiant and greater
than that of the X-ray source at the same time, and so reprocessing of
light from either is unlikely to dominate.  It therefore seems
unlikely that an accretion disc around the compact object can dominate
the optical brightening in outburst.

The relatively low temperature and large area inferred for the excess
light, if it originates in black body emission, can more plausibly be
associated with the equatorial material around the sgB[e] star than an
accretion disc around the compact object.  For a temperature of
$\sim10\,000$\,K, the extra emission requires an area $\sim40\times$
that of the sgB[e] star.  Both the temperature and size are plausible
for the equatorial material around a sgB[e] star.  In this case the
timescale arguments are similar to those for the lines.  The optical
decay appears slower than the X-rays (e.g.\ Clark et al.\
\cite{Clark:2000a}).  The peak optical luminosity of the continuum
source is also large; for $T\sim10\,000$\,K and an area $\sim40\times$
that of the star, the luminosity was at least
$\sim2\times10^5$\,L$_{\odot}\sim8\times10^{38}$\,erg\,s$^{-1}$
(assuming $d\sim5$\,kpc).  This is actually larger than the peak X-ray
luminosity at this distance, $L_{\rm
X}\sim3\times10^{38}$\,erg\,s$^{-1}$, and by the time of the optical
observations, the X-ray flux had already dropped by more than an order
of magnitude (Belloni et al.\ \cite{Belloni:1999a}).  Both the decay
timescale and optical luminosity render it unlikely that the continuum
brightening can come from X-ray reprocessing.  The timescales would be
more plausible for an interaction of the equatorial material with the
radio ejecta; assuming an expansion speed $\sim5000$\,km\,s$^{-1}$,
this area will be covered in less than half a day, so a rapid optical
brightening is possible in this way, i.e.\ powered by the kinetic
energy of the ejecta.

The remaining possibility is that there is a considerable direct
optical contribution from the ejecta from the explosion.  The
similarity of timescales of the optical and radio decay (Clark et al.\
\cite{Clark:2000a}) argue for this interpretation.  Optical
synchrotron can probably be ruled out.  The dereddened magnitude of
the additional continuum component of $V \la 6$ corresponds to a flux
of $\ga15$\,Jy early in the outburst.  Contemporaneous radio
observations were much weaker than this, with the spectral energy
distribution peaking at 650\,mJy at 8.4\,GHz and decreasing at higher
frequencies (Hjellming \& Mioduszewski \cite{Hjellming:1998b}).
Optical synchrotron emission thus seems unlikely without a very
unusual electron energy distribution.  However, if the accretion rate
was highly supercritical, we might expect much of the energy released
to be reprocessed by a scattering envelope to produce a bright optical
source (Shakura \& Sunyaev \cite{Shakura:1973a}).  Such a
supercritical accretion scenario has been invoked for a number of
X-ray binaries. \object{SS~433} is likely to a persistently
supercritical source (e.g.\ Fabrika \cite{Fabrika:1997a} and other
authors; see also Okuda \& Fujita \cite{Okuda:2000a} and references
therein).  \object{V4641~Sgr}, which also had an extremely short but
luminous X-ray outburst has been suggested to have undergone a
transient burst of supercritical accretion (Revnivtsev et al.\
\cite{Revnivtsev:2002a}).  Shakura \& Sunyaev (\cite{Shakura:1973a})
predict that the optical spectrum in the supercritical regime should
appear as a power-law, $F_{\nu} \propto \nu^{1/2}$, saturated with
broad emission lines.  We have already discussed the origin of the
broad emission lines that we see and suggested that these may be
related to outflowing material from supercritical accretion.  The
outburst optical continuum emission is clearly redder than the
underlying star, and the excess continuum emission is consistent with
a $F_{\nu} \propto \nu^{1/2}$ power law (Fig.~\ref{FluxSpecModelFig}).
This therefore does seem a plausible interpretation.  If this
supercritical accretion scenario is correct, then the observations
suggest that the accretion rate was extremely high and hence the
spherisation radius large; assuming a terminal velocity for the
ejected material of $\sim5000$\,km\,s$^{-1}$, and an optical
luminosity $\sim10^{38}$\,erg\,s$^{-1}$, suggests an accretion rate of
$\dot{M}\sim10^2$--$10^4$\,$\dot{\rm M}_{\rm crit}$ (Shakura \&
Sunyaev \cite{Shakura:1973a}), consistent with our estimate above of
the possible periastron accretion rate.

All of these calculations are extremely simplistic.  Our aim is not to
present a detailed spectral model, but rather to test which
explanations of the outburst emission are plausible.  X-ray heating,
whether of the sgB[e] star itself, the equatorial material, an
accretion disc around the compact object or extended dust, is not
consistent with observations.  Heating of the equatorial material by
the interaction with the expanding radio remnant, or direct emission
from these ejecta, remain possibilities.  The latter has the advantage
that invoking a supercritical accretion regime explains both the
optical continuum emission and the broad, blue-shifted emission
components, so is our favoured interpretation, but it is likely that a
combination of these mechanisms is involved.

\subsection{Is \object{CI~Cam} an X-ray binary?}

Finally we emphasise that none of the preceding discussion depends on
\object{CI~Cam} being a binary system.  We, and others, have argued
that \object{CI~Cam} is an sgB[e] star, and a compact object of some
kind is clearly implicated in the X-ray outburst, but it is not
necessary for them to be physically associated.  Indeed, the
properties of \object{CI~Cam} in quiescence are sufficiently similar
to other hot, luminous sgB[e] stars that if it is a binary then the
compact object's influence appears to only be important during an
outburst.  For example, it clearly does not truncate the disc as can
happen in classical Be X-ray binaries.  While it could be that
\object{CI~Cam} does have a compact object in a long period orbit, it
is also possible that the outburst could have resulted from a chance
encounter of an isolated black hole or neutron star with the
circumstellar material.  As a massive, young object, \object{CI~Cam}
is likely located in a region of recent star formation, so there are
likely to be many stellar remnants in its proximity.  While such a
chance encounter still seems improbable, it cannot be disproved based
on the statistics of a single event.  So far we have no conclusive
evidence for the binarity of \object{CI~Cam}.  Barsukova et al.\
(\cite{Barsukova:2002a}) have suggested that two periodicities are
present.  Their short 11.7\,day period is certainly too small to
represent the orbital period in a system such as this as the compact
object would have to be almost on the surface of the sgB[e] star, or
even inside it.  The longer 1100\,day period would be more plausible,
but cannot be considered convincing until it is seen to repeat for
multiple cycles.  In balance, while the hypothesis that
\object{CI~Cam} is a binary seems most likely, it is not proven and
the alternative, that there was a chance encounter with an isolated
compact object, cannot be ruled out.
%
%%%%%%%%%%%%%%%%%%%%%%%%%%%%%%%%%%%%%%%%%%%%%%%%%%%%%%%%%%%%%%%%%%%%%%%%%%%%%%%%%
%
\section{Summary and conclusions}
\label{conc}
\object{CI~Cam} is an sgB[e] star, and many of its characteristics,
particularly the emission line spectrum, are representative of this
class.  Comparison with other sgB[e] stars suggests that
\object{CI~Cam} is among the hottest members of the class and viewed
nearly pole-on.  It differs from the other sgB[e] stars in interacting
with a compact star which introduces an additional variable element.
It seems most likely that the compact object is physically associated
with \object{CI~Cam}, i.e.\ that it is an HMXB.  In this case, the
compact object is probably in a long period, eccentric orbit.  However
we cannot rule out the possibility that this was a chance encounter
and that the two stars are not physically associated.  Resolution of
this issue will likely involve waiting for a true periodicity,
repeated over several cycles, or another outburst.  In considering the
outburst mechanism, there is actually little difference between a long
period eccentric orbit and a chance encounter, so our discussion of
the outburst applies to both cases.

We suggest that the majority of the optical emission lines originate
from an equatorially concentrated outflow or circumstellar disc.
During outburst, hydrogen and helium lines appear to have two
components, a narrow rest component and a moderately blueshifted broad
component.  Metallic lines are mainly dominated by the narrow
component at all times, although some asymmetry is seen in outburst
suggesting that a broad component is present.  The square profiles of
the Fe\,\textsc{ii} lines can be explained by the equatorial outflow
model, if viewed pole-on, so the narrow component is likely associated
with this.  The broad component becomes weaker and narrower on the
decline, and almost disappears in quiescence.  This may come from
material ejected in the outburst rather than from the equatorial
outflow.  Forbidden lines fall into two categories; [O\,\textsc{i}]
and [Fe\,\textsc{ii}] show similar line profiles to Fe\,\textsc{ii},
but with a lack of low velocity material.  These profiles could
originate from the low density, upper layers of the equatorial
outflow.  Other forbidden lines, [N\,\textsc{ii}] and
[O\,\textsc{iii}] are narrower and likely come from a much more
extended region.

The outburst mechanism remains undetermined, although the outburst was
probably precipitated by the passage of the compact object through the
equatorial material.  It is unlikely that X-ray heating of any
component is responsible for the optical outburst.  Instead the
optical outburst is likely associated with the expanding remnant
produced by the X-ray outburst, either through direct emission from
the remnant or as a result of its interaction with the circumstellar
material.  The spectral shape of the outburst optical continuum, and
the presence of broad, blue-shifted emission components, are both
consistent with predictions for supercritical accretion resulting in
ejection of much of the material (Shakura \& Sunyaev
\cite{Shakura:1973a}), and the peak mass transfer rate for an
equatorial passage of the compact object is indeed predicted to be
well above the Eddington limit.

After the outburst changes in the emission lines persist for at least
three years, with Fe\,\textsc{ii} lines stronger than before and
He\,\textsc{i}, He\,\textsc{ii}, and N\,\textsc{ii} lines weaker.  The
timescale for the extended Fe\,\textsc{ii} decay, at least, is similar
to the expected viscous timescale of the disc of hundreds to thousands
of days, so this may indicate the gradual recovery of the disc to its
equilibrium state. As the system does not yet appear to have
stabilised continued monitoring is important to determine if the
system eventually recovers to the pre-outburst state or if it settles
to a different level.
\begin{acknowledgements}
We would like to thank Simon Jeffrey, Amy Mioduszewski, Guy Pooley,
John Porter, Rob Robinson, and Lev Titarchuk for information and many
helpful thoughts and discussions which have helped us converge on the
picture, albeit incomplete, that we now have of \object{CI~Cam}.  RIH
would particularly like to thank Rob Robinson for access to an
annotated high resolution spectrum of \object{CI~Cam} which proved
invaluable in identifying lines, and for permission to reproduce the
data shown in Fig.~\ref{TitaniumFig}b.

RIH, PAC, and CAH acknowledge support from grant F/00-180/A from the
Leverhulme Trust.  EAB and SNF acknowledge support from Russian RFBR
grant N\,00-02-16588.  MRG acknowledges the support of NASA/LTSA grant
NAG5-10889.  PR acknowledges support via the European Union Training
and Mobility of Researchers Network Grant ERBFMRX/CT98/0195.  WFW was
supported in part by the NSF through grant AST-9731416.

The William Herschel Telescope is operated on the island of La Palma
by the Isaac Newton Group in the Spanish Observatorio del Roque de los
Muchachos of the Instituto de Astrof\'\i{}sica de Canarias.  The
G. D. Cassini telescope is operated at the Loiano Observatory by the
Osservatorio Astronomico di Bologna.  Skinakas Observatory is a
collaborative project of the University of Crete, the Foundation for
Research and Technology-Hellas and the Max-Planck-Institut f\"{u}r
Extraterrestrische Physik.  This work also uses archival observations
made at Observatoire de Haute Provence (CNRS), France.  We would like
to thank K. Belle, P. Berlind, N. V. Borisov, M. Calkins, A. Marco,
D. N. Monin, S. A. Pustilnik, H. Quaintrell, T. A. Sheikina,
J. M. Torrej\'{o}n, A. V. Ugryumov, G. G. Valyavin, and R. M. Wagner
for assistance with some of the observations.

This work has made use of the NASA Astrophysics Data System Abstract
Service, the Vienna Atomic Line Database (Kupka et al.\
\cite{Kupka:1999a}) and Peter van Hoof's Atomic Line List v2.04
(http://www.pa.uky.edu/\-$\sim$peter/\-atomic/).

\end{acknowledgements}

\end{document}